\pdfoutput=1  

\documentclass[aps,prd,superscriptaddress,floatfix,nofootinbib,preprintnumbers,eqsecnum,twocolumn]{revtex4-1}
\usepackage[utf8]{inputenc}
\usepackage{graphicx}
\usepackage{amssymb,amsmath,multirow}
\usepackage{comment,color,placeins,centernot,amsmath,amsfonts,enumerate}

\usepackage{hyperref}
\usepackage[usenames,dvipsnames]{xcolor}
\hypersetup
{
  colorlinks,
  linkcolor={blue!80!black},
  citecolor={blue!70!black},
  urlcolor={blue!70!black}
}

\makeatletter
\renewcommand*\env@matrix[1][\arraystretch]{%
  \edef\arraystretch{#1}%
  \hskip -\arraycolsep
  \let\@ifnextchar\new@ifnextchar
  \array{*\c@MaxMatrixCols c}}
\makeatother

\def\gsim{\gtrsim}
\def\lsim{\lesssim}

\newcommand{\alignedfrac}[2]{%
    \setbox0\hbox{$#1$}        
    \dimen0=\wd0               
    \setbox1\hbox{$#2$}        
    \dimen1=\wd1               
    \ifdim\wd0<\wd1            
        \dfrac{#1\hfill}{#2}   
    \else                      
        \dfrac{#1}{#2\hfill}   
    \fi
}

\def\beq{\begin{equation}}
\def\eeq{\end{equation}}
\def\beqn{\begin{eqnarray}}
\def\eeqn{\end{eqnarray}}

\def\medwhitestar{\text{\fontsize{4.5}{8}\selectfont\hspace{-0.18mm}$\bigstar$\hspace{-0.23mm}}}
\def\medwhitestarcap{\text{\fontsize{4.0}{8}\selectfont\hspace{-0.18mm}$\bigstar$\hspace{-0.24mm}}}

\def\calM{{\cal M}}

\def\IR{\relax{\rm I\kern-.18em R}}
 \font\cmss=cmss10 \font\cmsss=cmss10 at 7pt
\def\IQ{\relax{\rm I\kern-.18em Q}}
\def\IZ{\relax\ifmmode\mathchoice
 {\hbox{\cmss Z\kern-.4em Z}}{\hbox{\cmss Z\kern-.4em Z}}
 {\lower.9pt\hbox{\cmsss Z\kern-.4em Z}}
 {\lower1.2pt\hbox{\cmsss Z\kern-.4em Z}}\else{\cmss Z\kern-.4em Z}\fi}

\newcommand{\nn}{\nonumber}
\newcommand{\abs}[1]{\left| #1 \right|}

\newcommand{\p}{\partial}
\newcommand{\ve}[1]{\mathbf{#1}}

\newcommand{\il}[1]{\mbox{$#1$}} 

\newcommand{\Veff}{V_{\rm eff}}
\newcommand{\phidot}{\dot{\phi}}

\newcommand{\tp}{t_{\rm p}}
\newcommand{\Nkin}{N_{\rm kin}}

\def\ie{{\it i.e.}\/}
\def\eg{{\it e.g.}\/}

\def\mbar{\overline{m}}
\def\msum{m_{\rm sum}}
\def\thetabar{\overline{\theta}}
\def\thetadot{\dot{\theta}}
\def\thetaddot{\ddot{\theta}}
\def\rhobar{\overline{\rho}}

\def\rhoR{\rho_{R}}
\def\rhoS{\rho_{S}}
\def\wphi{w_{\phi}}
\def\lambdabar{\overline{\lambda}}

\def\alphabar{\overline{\alpha}}
\def\betabar{\overline{\beta}}

\def\phiLN{\phi_{\lambda_0}}
\def\phiLNI{\phi_{\lambda_0}^{{\color{white}{X_2}}}{\hspace{-10.0pt}\raisebox{3.6pt}{\scriptsize (\rm I)}}}
\def\phiLNE{\phi_{\lambda_0}^{{\color{white}{X_2}}}{\hspace{-10.0pt}\raisebox{3.6pt}{\scriptsize (\rm E)}}}
\def\phiLNIexp{\phi_{\lambda_0}^{{\color{white}{X_2}}}{\hspace{-10.0pt}\raisebox{3.6pt}{\tiny (\rm I)}}}
\def\phiLNEexp{\phi_{\lambda_0}^{{\color{white}{X_2}}}{\hspace{-10.0pt}\raisebox{3.6pt}{\tiny (\rm E)}}}

\def\varphihat{\widehat{\varphi}}
\def\Ninf{N_{\rm inf}}
\def\meff{m_{\rm eff}}
\def\Aphi{\mathcal{A}_{\phi}}
\def\Q{\mathcal{Q}}
\def\half{{\textstyle \frac{1}{2}}}

\newcommand{\e}{$e$}

\newcommand{\Mpc}{\text{Mpc}}

\begin{document}

\title{Enlarging the Space of Viable Inflation Models:\\ A Slingshot Mechanism}
\author{Keith R. Dienes}
\email{dienes@email.arizona.edu}
\affiliation{Department of Physics, University of Arizona, Tucson, AZ 85721 USA}
\affiliation{Department of Physics, University of Maryland, College Park, MD 20742 USA}
\author{Jeff Kost}
\email{jeffkost@ibs.re.kr}
\affiliation{\mbox{Center for Theoretical Physics of the Universe, Institute for Basic Science, Daejeon 34126 Korea}}
\author{Brooks Thomas}
\email{thomasbd@lafayette.edu}
\affiliation{Department of Physics, Lafayette College, Easton, PA 18042 USA}

\preprint{CTPU-PTC-19-22}

\begin{abstract}
    The viability of a given model for inflation is determined not only by the form of the 
    inflaton potential, but also by the initial inflaton field configuration.  In many models, 
    field configurations which are otherwise well-motivated nevertheless fail to induce inflation, or fail to 
    produce an inflationary epoch of duration sufficient to solve the horizon and flatness problems.   
    In this paper, we propose a mechanism which enables inflation to occur even with such
     initial conditions.  
    Our mechanism involves multiple scalar fields which experience a time-dependent mixing.  
    This in turn leads to a ``re-overdamping'' phase as well as a parametric resonance 
    which together ``slingshot'' the inflaton field from regions of parameter space that do not induce inflation
    to regions that do.  Our mechanism is flexible, dynamical, and capable of yielding
    an inflationary epoch of sufficiently long duration.  This slingshot mechanism can therefore be 
    utilized in a variety of settings and thereby enlarge the space of potentially viable inflation models.
\end{abstract}

\maketitle




\section{Introduction, motivation,\\ and summary\label{sec:Intro}}


Because of its many properties, both theoretical and observational,
the inflationary paradigm has become a standard component of early-universe 
\mbox{cosmology~{\mbox{\cite{Starobinsky:1980te,Guth:1980zm,Linde:1981mu,Albrecht:1982wi,Olive:1989nu,Baumann:2009ds}}}}.
Indeed, many observations and cosmological fine-tuning problems --- {\it e.g.}\/, the horizon problem, 
the absence of primordial topological defects, the flatness problem --- can be understood by positing 
an epoch of rapid inflationary expansion in the early universe.
Moreover, models for inflation generally predict a nearly scale-invariant 
spectrum of primordial perturbations, in agreement with observations of the 
cosmic microwave background (CMB) and large-scale 
\mbox{structure~{\mbox{\cite{Press:1980zz,Mukhanov:1981xt,Hawking:1982cz,Starobinsky:1982ee,Guth:1982ec,Bardeen:1983qw}.}}}

One critical ingredient in any inflationary scenario is an appropriate scalar
potential $V(\phi)$ for the inflaton $\phi$.  The form of $V(\phi)$ is constrained
by observations --- in particular, inflation typically only occurs when 
$\phi$ traverses a sufficiently flat region of the potential.
Likewise, an inflationary model also requires the specification of
a set of initial conditions for the inflaton field within
the phase space \il{\{\phi,\phidot\}}.
Only certain regions within this phase space 
give rise to inflationary dynamics, and only certain subregions thereof
lead to an inflationary epoch of sufficient duration.  
Indeed, the number of \e-folds required for successful inflation generally 
falls within a range \il{\Ninf \gtrsim 50\text{--}60}, where the uncertainty is due to our 
ignorance of the amount of expansion which occurs during the 
reheating epoch~\cite{Liddle:2003as}.
As a result, the initial field
configuration \il{\{\phi,\phidot\}}  is another critical ingredient in 
determining the phenomenological viability of a given inflation model.

\begin{figure}[t]
    \begin{center}
        \includegraphics[keepaspectratio, width=0.499\textwidth]{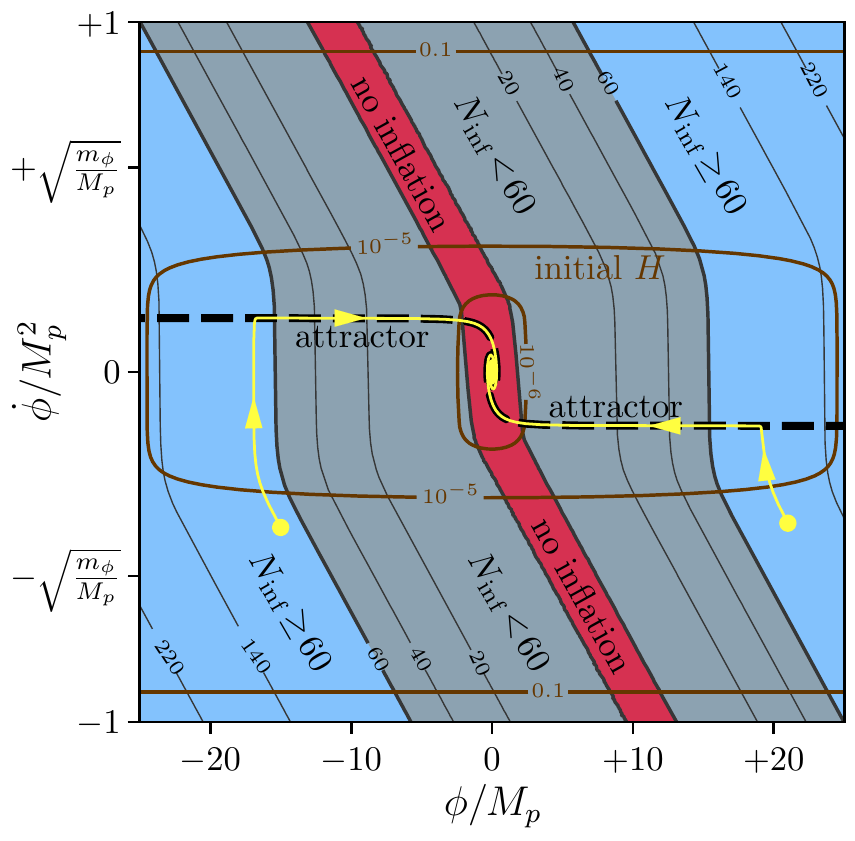}
    \end{center}
    \caption{The space of possible initial configurations \il{\{\phi,\phidot\}}
    for the inflaton field in the case of a quadratic potential \il{V(\phi) = \half m_{\phi}^2\phi^2} 
    with \il{m_\phi = 10^{-6}M_p}. 
    The blue regions indicate the locations that can yield an inflationary epoch of 
    duration (here taken to be \il{\gtrsim\mathcal{O}(60)}~\e-folds) sufficient to solve the 
     horizon and flatness problems,
    while the gray regions lead to insufficient inflation
    and the red regions lead to no inflation at all.  The brown contours indicate
    the initial Hubble scale $H_{\rm I}$ for each initial configuration \il{\{\phi,\phidot\}},
    and two example trajectories (yellow) for the inflaton field are also shown. 
    The black dashed curves show the attractors along which inflation occurs.}
\label{fig:intro_phase_space}
\end{figure}

As an example, let us consider perhaps the simplest inflaton potential, 
\il{V(\phi) = {\textstyle {1\over 2}} m_\phi^2 \phi^2}, where $m_{\phi}$ is the inflaton mass.
In Fig.~\ref{fig:intro_phase_space} we show 
the corresponding phase space \il{\{\phi,\phidot\}}, 
with contour lines indicating the number of $e$-folds
of inflation that are eventually produced
if our field begins at that location.  
We have also colored the different 
regions of this phase space accordingly, 
so that initial configurations within the red region do not lead to inflation at all
while points within the gray region lead to insufficient
inflation (defined for this figure as having fewer than 60 $e$-folds).
Only the regions in blue lead to an inflationary epoch of
sufficient duration.   

These results can be understood in terms of the dynamical
flow of the inflaton field within this phase space.
In general, regardless of its initial location,
the field will travel 
along an $e$-fold contour line in the direction of one of the 
black-dashed attractor curves at \il{\phidot = \pm \sqrt{2/3}\, m_{\phi}M_p}
where $M_p$ is the reduced Planck mass.
Two sample flows are indicated in yellow in Fig.~\ref{fig:intro_phase_space}.~
Only when the inflaton reaches the attractor does inflation begin,
and only when the inflaton joins the attractor at sufficiently
large $|\phi|$ will this inflationary epoch have sufficient duration.
Thus, as a result of this dynamics, we see that the regions of phase space
which lead to sufficiently long inflationary epochs
are those with sufficiently large initial values of $|\phi|$ or $|\phidot|$ or both.
Similar conclusions hold for a wide variety of potentials.

Within the context of such potentials, many different scenarios 
exist for generating such initial conditions involving large $|\phi|$ or $|\phidot|$.
For example, in chaotic inflation~\cite{Linde:1983gd},   
initial conditions are assumed to be
randomly distributed, subject to a Planckian bound
on the total energy density \il{\rho \leq M_p^4}.
One then finds that for small $m_\phi$, the initial vacuum expectation values (VEVs) for
$|\phi|$ and/or $|\phidot|$ are typically large, the former even trans-Planckian.
Likewise, natural inflation~{\mbox{\cite{Freese:1990rb,Freese:2004un}}}
also implements such a large field VEV for $|\phi|$, even if $|\phidot|$ in 
such scenarios is typically small.

Our focus here, by contrast, is on the opposite situation:
what happens if both $|\phi|$ and $|\phidot|$ are initially small, perhaps
both even sub-Planckian?  
After all, sub-Planckian field VEVs might be viewed as more natural from an effective field theory point of view.
However, in many inflationary models, such initial field VEVs correspond 
to regions of phase space in which slow-roll conditions will be severely violated and inflation will not occur.
Is there any way inflation can be salvaged in such scenarios?

In this paper, we shall propose a mechanism which accomplishes precisely this. 
We shall refer to this as a ``slingshot'' mechanism:  even if 
we begin within a region of phase space from which inflation would not 
ordinarily be expected to occur, the dynamics of the system 
can ``slingshot'' the inflaton field into a different region of 
phase space in which it does.   Our slingshot mechanism is built upon 
a brief time-dependent modification to the scalar potential in 
the inflationary sector of the theory --- precisely as might 
arise as the consequence of a pre-inflationary cosmological 
phase transition.   As we shall discuss, the resulting inflaton 
dynamics includes a short-lived but robust parametric resonance 
followed by a unique re-overdamped phase, and together these two features 
propel the inflaton field into new regions of phase space which 
potentially exhibit the desired large field VEVs.   Indeed,  we 
shall find that both of these features generically arise together 
in such scenarios.    Our slingshot mechanism is therefore 
completely flexible, dynamical, and capable of giving rise to 
an inflationary epoch of sufficiently long duration.  We also 
expect that our mechanism is also rather general, and can be 
utilized in a variety of different settings and for a variety of 
different potentials. This mechanism can thereby enlarge 
the space of potentially viable inflation models.

This paper is organized as follows.   We begin in Sect.~\ref{sec:TheScalarSector}
by presenting the physics of our slingshot mechanism.  
As we shall discuss, the natural arena for this physics
is a sector of scalar fields experiencing a mass-generating phase transition.
Specifically, we shall focus our study on the simple yet illustrative example of a 
sector consisting of two real scalar fields.
Then, in Sect.~\ref{sec:Inflation}, we embed this sector within the context
of a general cosmology and demonstrate that our slingshot mechanism can indeed
give rise to inflation --- even when our initial field configuration is within
otherwise ``forbidden'' regions such as the red region in Fig.~\ref{fig:intro_phase_space}.~
We investigate the general parameter space associated
with our mechanism and determine those regions in which sufficient inflation can be produced.
In this way we demonstrate that our mechanism can give rise to not only an inflationary
epoch but also one with a number of $e$-folds sufficient for phenomenological purposes.
In Sect.~\ref{sec:Mapping}, we then develop an analytic approach to understanding the numerical
results presented in Sect.~\ref{sec:Inflation} --- an approach which allows us 
to forge a direct connection between
the underlying parameters which characterize the phase transition and the relevant inflationary
parameters and observables.  In Sect.~\ref{sec:Constraints} we then exploit this connection in
order to establish constraints on this minimal realization of our ``slingshot'' mechanism.
Finally, in Sect.~\ref{sec:Conclusions}, we summarize our main results, elaborate on the
various implications of our slingshot mechanism, and discuss a number of potential extensions
and generalizations of the minimal scenario we have presented here.
In Appendix~\ref{sec:ParticleProduction}, we discuss constraints that
can arise from particle production due to non-adiabatic changes in the vacuum state.

\section{The scalar sector and the slingshot mechanism\label{sec:TheScalarSector}}


As discussed in Sect.~\ref{sec:Intro},
our slingshot mechanism is built upon the 
non-trivial dynamics of a two-scalar system which undergoes a mass-generating phase transition.
In this section we shall therefore introduce our scalar system
and discuss its dynamics, with the 
goal of understanding how and why a slingshot emerges.

\subsection{Two scalars and a mass-generating phase transition}


We begin by assuming the existence of a non-minimal scalar sector which 
experiences a mass-generating phase transition in the early universe, prior to
an inflationary epoch. In particular, for concreteness we shall imagine
that this sector consists of two scalar fields \il{\lbrace \phi_0,\phi_1\rbrace}
whose dynamics is governed by an effective potential $\Veff(\phi_0,\phi_1,t)$ 
which depends not only on the field VEVs $\phi_0$ and $\phi_1$ (thereby \emph{implicitly} introducing
a time-dependence for $\Veff$ as our two fields evolve in field space) but also on other 
parameters which may carry their own \emph{explicit} time-dependence as the result of 
a possible phase transition.  In a spatially flat Friedmann-Robertson-Walker (FRW) cosmology, 
the corresponding action for these fields then takes the form
\beq
\mathcal{S}_{\phi} =  \int d^4x \sqrt{-g}\biggl[
   \half   (\partial_\mu \phi_0)^2 
   + \half  (\partial_\mu \phi_1)^2 
            - V_{\rm eff}(\phi_0,\phi_1,t)\biggr] 
\label{eq:S_phi}
\eeq
where $g$ is the determinant of the FRW metric.
While many algebraic forms for $V_{\rm eff}(\phi_i,t)$ are possible,
for simplicity we shall henceforth assume that the contributions 
to $\Veff(\phi_i,t)$ are of the form
\beq
   V_{\rm eff}(\phi_i,t) ~=~ {1\over 2} \sum_{i,j}\phi_i\mathcal{M}^2_{ij}(t)\phi_{j}~,
\label{eq:Veff}
\eeq
where $\mathcal{M}^2(t)$ is the corresponding {\il{2\times 2}} time-dependent squared-mass
matrix for our two scalars.  Under this assumption, our cosmological phase transition
can be viewed as a \emph{mass-generating} phase transition.  As we shall see, a sector
consisting of only two scalar fields is sufficiently complex to give rise to all of the
phenomena which shall eventually interest us, but also sufficiently simple that
our analysis remains relatively straightforward and tractable.  We emphasize, however, that
there will be nothing in our eventual results that requires that there be only two fields in
this scalar sector.  Indeed similar phenomena can emerge in more complicated scenarios.

In order to maintain as much generality as possible, we shall follow 
Refs.~{\mbox{\cite{Dienes:2015bka,Dienes:2016zfr}}} in parametrizing the effects 
of our phase transition on the mass matrix $\calM^2$.  First, we shall let $t_G$ denote 
the time at which the phase transition occurs and assume that at early times \il{t\ll t_G} 
the mass matrix takes the simple form
\beq\label{eq:M2early}
\mathcal{M}^2 ~=~ 
\begin{bmatrix}
    0 & 0\\
    0 & M^2
\end{bmatrix} ~~~~~ {\rm for}~~ t\ll t_G~.
\eeq
In other words, we are assuming that our lighter field is initially massless 
while our heavier field has a non-zero mass $M$.  Next, we shall 
assume that our phase transition produces additional contributions to this mass matrix
such that for times \il{t \gg t_G} long after the phase transition 
our mass matrix takes the late-time asymptotic form
\beq
\label{eq:M2late}
\mathcal{M}^2  ~=~ 
\begin{bmatrix}[1.1]
    0 & 0\\
    0 & M^2
\end{bmatrix}
+
\begin{bmatrix}[1.1]
    \mbar_{00}^2 & \mbar_{01}^2\\
    \mbar_{01}^2 & \mbar_{11}^2
\end{bmatrix} ~~~~~ {\rm for}~~ t\gg t_G~,
\eeq
where $\mbar_{ij}^2$ represent the late-time contributions to the mass matrix arising from 
the phase transition.  In this way, we allow for the possibility that our phase transition 
not only modifies the masses of our states, but also introduces a non-trivial mixing between 
them.  Finally, we shall make the natural assumption that the phase transition unfolds smoothly 
in such a way that these extra contributions to $\calM^2$ are all generated with a uniform 
time-dependence.  In other words, for any arbitrary time $t$, we shall assume that the mass 
matrix takes the form
\beq\label{eq:M2def}
\mathcal{M}^2(t) ~=~
\begin{bmatrix}[1.1]
    0 & 0\\
    0 & M^2
\end{bmatrix}
+
\begin{bmatrix}[1.1]
    m_{00}^2(t) & m_{01}^2(t)\\
    m_{01}^2(t) & m_{11}^2(t)
\end{bmatrix}~, 
\eeq
where we can write
\beq
  m_{ij}(t) ~=~ \mbar_{ij}\cdot h(t)
  \label{eq:mijtimedep}
\eeq
with $h(t)$ representing a monotonic function $h(t)$ such that
\il{h(t)\to 0} as \il{t/t_G\to 0} and \il{h(t) \to 1} as \il{t/t_G\to \infty}.
This function $h(t)$ completely characterizes the manner in which the phase 
transition unfolds as a function of $t$.  
We shall enforce our expectation that $t_G$ is the
``central time'' of the phase transition by defining $t_G$
through the relation \il{h(t_G) =1/2}.  Furthermore, the most significant
changes to $h(t)$ occur during a window
of approximate duration $\Delta_G$ centered around $t_G$ 
--- \ie, within the time interval \il{[t_G-\Delta_G/2,t_G+\Delta_G/2]}.
Note that the mass matrix $\calM^2$ in Eq.~(\ref{eq:M2def}) 
is Hermitian and has non-negative eigenvalues
at any instant in time so long as all of the $m_{ij}^2$ are real, 
with \il{m_{00}^2\geq 0}, \il{m_{11}^2 \geq -M^2}, and 
\il{m_{00}^2 (M^2 + m_{11}^2) \geq m_{01}^4}.

Beyond these requirements, most of the qualitative results of this paper are
largely independent of the particular choice of functional form for $h(t)$.  
Indeed, for many purposes we may regard $h(t)$ rather than $t$ itself as our cosmological 
clock variable, choosing to study the dynamics of our system relative to our ``$h$-clock''
without worrying about the particular mapping between $h(t)$ and the true cosmological time $t$.
However, when necessary, we shall adopt the specific choice~{\mbox{\cite{Dienes:2015bka,Dienes:2016zfr}}}
\beq
  h(t) ~=~ 
  {1\over 2} \left\{1 + \text{erf}\left[\frac{\sqrt{\pi}t_G}{\Delta_G}
  \log\left(\frac{t}{t_G}\right)\right]\right\} ~,
  \label{hdef}
\eeq
where the parameter $\Delta_G$, as discussed above, represents the approximate time interval 
over which the phase transition unfolds.  Indeed, as discussed in Ref.~\cite{Dienes:2015bka},
the algebraic form in Eq.~(\ref{hdef}) satisfies all of our requirements and provides a 
compelling model for a generic phase transition, provided that we take 
\il{\Delta_G\leq \sqrt{2\pi} t_G}.

Given the action in Eq.~(\ref{eq:S_phi}),
our fields $\phi_{0,1}$ then evolve according to the equations of motion
\beq
  \ddot{\phi}_i + 3H\phidot_i + \sum_j\mathcal{M}^2_{ij}\phi_j ~=~ 0~,
  \label{eq:eqnsofmotion}
\eeq
where $H(t)$ is the Hubble parameter.
With the mass matrix given in 
Eq.~(\ref{eq:M2def}),
the dynamics of the scalar sector is thus nearly completely determined.
Indeed, it remains only to specify
initial conditions for our fields at some initial time $t_{\medwhitestar}$ prior to the 
phase transition.  

Given our ultimate goal of
extending the space
of initial field configurations for inflation 
into regions that would {\it a priori}\/ 
experience no inflation at all,
a particularly convenient choice of initial conditions 
is to assume that only our initially 
massless field $\phi_0$ has a non-zero displacement,
\ie, that \il{\phi_i(t_{\medwhitestar}) = \Aphi\delta_{i0}}  where $\Aphi$ is a constant,
and that both fields start from rest, with \il{\dot \phi_i (t_{\medwhitestar}) = 0}. 
This choice of initial conditions ensures that our scalar sector has no energy prior the phase
transition, so that all energy in the scalar sector is derived
from the phase transition itself.  
Note that it is natural to take {\il{\phi_1(t_\medwhitestar)=\phidot_1(t_\medwhitestar)=0}} 
because the heavy mass $M$ of this field renders it initially underdamped, whereupon Hubble 
friction damps out any oscillations this field might undergo.  We can therefore assume that 
this is what has occurred prior to $t_\medwhitestar$.    However, there are also several reasons why we 
choose {\il{\phidot_0(t_\medwhitestar)=0}}.   First, even for a massless (and therefore overdamped) 
field $\phi_0$, Hubble friction damps out any initial field velocity $\phidot_0$ over 
a Hubble timescale, even while $\phi_0$ remains non-zero.    Moreover, we note that an initial 
condition with non-zero $\phi_0$ but zero $\phidot_0$ emerges naturally from production 
mechanisms such as misalignment production.   But most importantly, as explained 
in Sect.~\ref{sec:Intro}, our goal in this paper is to have our
initial values of $\phi_i$ and $\phidot_i$ both sub-Planckian.   
While our eventual results will prove somewhat sensitive to ${\cal A}_\phi$ (which is 
why we shall keep this as a free parameter throughout this paper), 
most of our results will turn out to be largely insensitive to the precise sub-Planckian value of 
$\phidot_0(t_\medwhitestar)$ as long as this value is of the same sign as ${\cal A}_\phi$.   
For all of these reasons, we shall assume 
that {\il{\dot\phi_0(t_\medwhitestar)=0}} in what follows.  
However, in Sect.~\ref{sec:Inflation} we shall return to this issue and 
demonstrate explicitly how our results would change 
if we allowed $\dot\phi_0(t_\medwhitestar)$ to vary. 

Our choice of the initial time $t_\medwhitestar$ also deserves comment.
Clearly we wish to choose a time which is much earlier than our phase transition, 
while our $h(t)$ function in Eq.~(\ref{hdef}) is still fairly close to zero.
We shall therefore define our fiducial time $t_\medwhitestar$ through 
the condition that {\il{h(t_\medwhitestar)=10^{-10}}},
and we shall use this value for $t_\medwhitestar$ for all explicit calculations in this paper. 
In particular, this choice is always larger than the Planck time,
and for values of $t_G$ and $\Delta_G$ within what will eventually be our main region of phenomenological
interest we find {\il{t_\medwhitestar \approx t_G - 2.3 \Delta_G}}.
However, the main qualitative results of this paper will be completely independent 
of $t_\medwhitestar$
so long as \il{t_\medwhitestar \ll t_G}. 

This, then, completely specifies the physics of the scalar sector.   
In particular, for any choice of the initial conditions $M$ and $\Aphi$, 
we see that there are five parameters which govern the resulting physics: 
two parameters $t_G$ and $\Delta_G$ which govern the temporal features of the phase transition,
and three parameters $\mbar_{00}^2$, $\mbar_{11}^2$, and $\mbar_{01}^2$ 
which describe the late-time contributions to the mass matrix that result from this phase transition. 
Because of the time-dependent nature of the mass matrix, our two fields $\phi_0$ and $\phi_1$
experience a non-trivial time evolution.  In particular, at any instant of time these 
two fields experience a mixing angle $\theta(t)$ given by
\beq 
     \tan(2\theta) ~\equiv ~ \frac{2 m_{01}^2}{M^2 - m_{00}^2 + m_{11}^2 }~.
\label{eq:theta}
\eeq
Thus, the mass eigenstates of our system  
$\phiLN$ and $\phi_{\lambda_1}$ --- along with 
their corresponding mass eigenvalues $\lambda_0$ and $\lambda_1$ --- are 
continually changing over the course of the phase transition.
At any instant, the mass eigenstate $\phi_{\lambda_i}(t)$ 
is overdamped if \il{\lambda_i (t) < 3H(t)/2} and otherwise underdamped.
Indeed it is only at late times \il{t\gg t_G} after the phase transition has passed that
our system asymptotes to one with particular fixed late-time mass eigenstates
$\phiLN$ and $\phi_{\lambda_1}$ with fixed late-time masses $\lambdabar_0$ and $\lambdabar_1$.
 
For our work, it shall prove convenient to 
recast the mass matrix (\ref{eq:M2def}) into the form
\beq\label{eq:massmatrixfull}
\mathcal{M}^2 ~=~
 \half \msum^2\sqrt{1-\alpha^2}
\begin{bmatrix}[1.1]
    \sqrt{\frac{1-\alpha}{1+\alpha}} & 1-\beta \\ 
    1-\beta & \sqrt{\frac{1+\alpha}{1-\alpha}}
\end{bmatrix} \ ,
\eeq
where we have defined the time-dependent quantities  
\beqn
    m_{\rm sum}^2 ~&\equiv &~  M^2 + m_{00}^2 + m_{11}^2 \nonumber\\
    \alpha ~&\equiv&~ \frac{ M^2 - m_{00}^2 + m_{11}^2}{ M^2 + m_{00}^2 + m_{11}^2} \nonumber\\
    \beta ~&\equiv&~ 1 - \frac{m_{01}^2}{\sqrt{m_{00}^2\left(M^2+m_{11}^2\right)}}~. 
\eeqn
Thus $\beta$ --- 
the only one of these quantities which depends on $m_{01}^2$ ---
parametrizes the degree to which our fields mix at any instant of time.
Requiring a Hermitian, positive-semidefinite mass matrix in Eq.~(\ref{eq:massmatrixfull})
then restricts us to the parameter range  \il{m_{\rm sum}^2 \geq 0}, \il{|\alpha|\leq 1}, and \il{0\leq \beta \leq 2},
with \il{\alpha=\pm 1} allowed only for \il{\beta = 1}.
Indeed, within these ranges, Eq.~(\ref{eq:theta}) now takes the form
\beq
  \tan (2\theta) ~=~ \frac{1-\beta}{\alpha}\sqrt{1-\alpha^2}~,
  \label{eq:theta2}
\eeq
where \il{-\pi\leq \theta \leq \pi}.
Likewise, at any moment in time, the eigenvalues of the mass matrix in
Eq.~(\ref{eq:massmatrixfull}) take the simple form
\beq
  \lambda_{0,1}^2 ~=~ 
  \half m_{\rm sum}^2\left[1 \mp \sqrt{\alpha^2 + (1-\alpha^2)(1-\beta)^2}\right]~,
  \label{eq:massspectrum}
\eeq
with \il{\lambda_0^2 + \lambda_1^2 = m_{\rm sum}^2}, as expected.
Since \il{\lambda_0\leq \lambda_1} in all cases, we see that $m_{\rm sum}^2/2$ is both
the maximum possible value for $\lambda_0^2$ and the minimum possible value for $\lambda_1^2$.

It is possible to exploit certain symmetries of this two-scalar model to restrict 
the ranges of our parameters still further.  For example, as long as we are only interested in understanding 
the behavior of our fields and their corresponding energy densities, our system is invariant 
under the transformation \il{(\phi_0,\phi_1)\to -(\phi_0,\phi_1)} in which the signs of both
fields are simultaneously flipped.  This symmetry allows us to restrict our attention to the range 
\il{-\pi/2 \leq \theta \leq \pi/2}.  Likewise, flipping the \emph{relative} signs 
of the fields --- \eg, taking \il{(\phi_0,\phi_1)\to (\phi_0,-\phi_1)} --- is tantamount to 
flipping \il{\beta \to 2-\beta}.  We can thus restrict our attention to values
of $\beta$ which lie within the range \il{0\leq \beta\leq 1} without loss of generality.
This corresponds to restricting \il{m_{01}^2\geq 0}, or equivalently restricting
our attention to \il{-\pi/4\leq \theta\leq \pi/4}, with positive (negative) values
of $\theta$ corresponding to positive (negative) values of $\alpha$.

Thus, for any values of $M$ and $\Aphi$, our scalar sector can be parametrized in terms
of $\mbar_{\rm sum}^2$, $\alphabar$, $\betabar$, $t_G$, and $\Delta_G$.   However, for any value of
$\mbar_{\rm sum}^2$ and $\alphabar$,  our result in Eq.~(\ref{eq:massspectrum}) provides 
a one-to-one relationship between $\betabar$ and the
late-time mass $\lambdabar_0$ of the lighter scalar $\phiLN$.
Since $\phiLN$ will eventually be identified with the inflaton,
and $\lambdabar_0$ with the inflaton mass,
we will occasionally trade $\betabar$ for $\lambdabar_0$ in what follows.

\FloatBarrier
\subsection{Building the slingshot:  Assembling the required ingredients}


In general, this two-field scalar system can exhibit 
a variety of different behaviors and thereby give rise to a rich set 
of possible phenomenologies~\cite{Dienes:2015bka}.
However, what shall concern us in this paper is the possibility of a ``slingshot'' effect in
which the VEV of the lighter field $\phiLN$ is propelled toward super-Planckian values.
We shall therefore henceforth focus on this possibility, and demonstrate that the
required ingredients emerge quite naturally within the setup we have described.

\begin{figure}[t!]
    \begin{center}
    \includegraphics[keepaspectratio, width=0.49\textwidth]{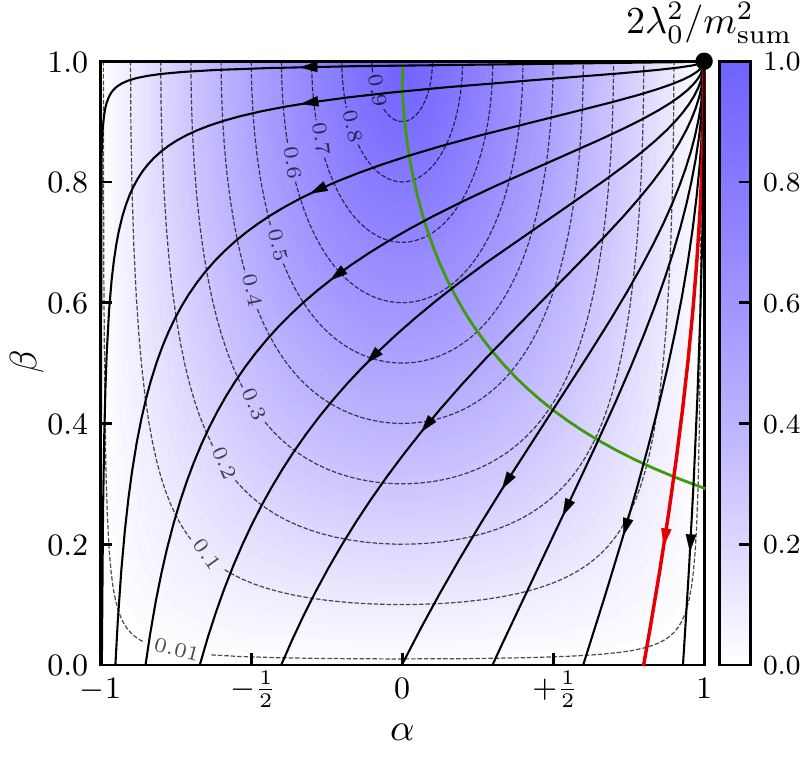}
    \end{center}
\vskip -0.2 truein
    \caption{A contour plot of the lighter eigenmass $\lambda_0^2$, normalized by its maximum possible 
       value \il{\msum^2/2}, across the \il{\{\alpha, \beta\}}  
      parameter space.  For this plot we have taken  \il{m_{\rm sum}^2 = M^2}, with \il{M= M_p/3}.
      Superimposed on this contour plot are ``flow lines'' (indicated with arrows), whose interpretations
        are discussed in the text.  
      When needed for concreteness later in this paper, we shall often consider the 
       ``benchmark'' flow line with \il{\alphabar=0.8} and \il{\betabar\ll 1} shown in red.
    An additional  green contour line connects the points along each flow line
       at which the value of $\lambda_0$ is maximized.}  
    \label{fig:lambda_contour}
\end{figure}

In order to understand how this dynamics emerges, let us begin by considering the behavior of the lighter mass
eigenvalue $\lambda_0$ as a function of the three parameters \il{\lbrace m_{\rm sum}^2, \alpha,\beta\rbrace}.  
The value of $\lambda_0$ is given in Eq.~(\ref{eq:massspectrum}), with contours of constant $\lambda_0^2$ plotted 
in Fig.~\ref{fig:lambda_contour} as fractions of its maximum value $m_{\rm sum}^2/2$. 
For this figure we have taken \il{m_{\rm sum}^2 = M^2} and plotted contours within the remaining
\il{(\alpha,\beta)} parameter space.

Superimposed on this contour plot we have also indicated various ``flow'' lines, denoted with arrows.
These lines indicate the time evolution of the system, and may be understood 
as follows.   Long before the phase transition,
our mass matrix takes the form given in Eq.~(\ref{eq:M2early}), corresponding
to \il{m_{\rm sum}^2= M^2} and \il{\alpha=\beta=1}.
This is therefore the initial point for our time flow.
In general, our system then evolves through different values ($m_{\rm sum}^2$, $\alpha$, $\beta$) 
as functions of time, and the particular path that is followed through this parameter space   
ultimately depends on the values of the chosen late-time parameters $\mbar_{ij}^2$ 
[or equivalently \il{(\mbar_{\rm sum}^2, \alphabar, \betabar)}] which parametrize
the particular phase transition under study.
However, for the special case \il{\mbar_{\rm sum}^2 = M^2} 
(or equivalently \il{\mbar_{11}^2 = -\mbar_{00}^2}), it turns out that $m_{\rm sum}^2$
remains fixed at $M^2$ for all times.
In this case our system evolves only within the two-dimensional 
$(\alpha,\beta)$ plane shown in Fig.~\ref{fig:lambda_contour}.~
Thus, for \il{\mbar_{\rm sum}^2=M^2} and any chosen \il{(\alphabar,\betabar)},
the flow line connecting \il{(\alpha,\beta)=(1,1)} to \il{(\alpha,\beta)=(\alphabar,\betabar)} 
indicates the path for the flow of our 
system as a function of time, with the corresponding values 
of $\lambda_0^2$ varying as the different $\lambda_0^2$-contour lines are crossed. 
This flow then terminates once the final location \il{(\alphabar,\betabar)} is reached.\footnote{
As an aside, we note that since a given flow line is specified by the 
choice of its final location \il{(\alphabar,\betabar)},
our ability to plot general flow lines without endpoints as in Fig.~\ref{fig:lambda_contour} 
rests upon a highly non-trivial fact:  any flow line passing through a given 
point \il{(\alpha^\ast,\beta^\ast)}
en route to its final location \il{(\alphabar,\betabar)} 
entirely subsumes the (\emph{a priori} distinct) 
flow line for which the final location is instead taken to be \il{(\alpha^\ast,\beta^\ast)}.
In other words, all flow lines which begin at \il{(\alpha,\beta)=(1,1)} and include a given intermediate 
point \il{(\alpha^\ast,\beta^\ast)} actually coincide for all portions of the flow between these two points
regardless of the particular time 
at which the \il{(\alpha^\ast,\beta^\ast)} point is reached.
Interestingly, this in turn implies that
any arbitrary shift \il{t^\ast\to t^\ast + \delta t} in the time $t^\ast$
at which a given intermediate point is reached
can be realized simply by shifting the final location
\il{(\alphabar,\betabar)} of the flow line
(and thereby shifting
the underlying theory).   
Moreover, this shift in 
\il{(\alphabar,\betabar)} 
simply moves \il{(\alphabar,\betabar)} 
along the same flow line.
In this sense, then, 
these flow lines --- now viewed as trajectories within
the \il{(\alphabar,\betabar)} plane --- have 
an additional interpretation as classical \emph{renormalization-group} flows which describe 
how our different \emph{theories} 
are connected under such time shifts.  While this discussion has been limited
to the special case with \il{m_{\rm sum}^2= M^2}, as shown
in Fig.~\ref{fig:lambda_contour},
these properties hold even 
for flows that traverse the full 
three-dimensional \il{(m_{\rm sum}^2,\alpha,\beta)} parameter space.}

For any specified final location \il{(\alphabar,\betabar)}, the results in 
Fig.~\ref{fig:lambda_contour} indicate how the lightest eigenvalue $\lambda_0$ 
varies as our system evolves in time.  For our purposes, however, the most 
important feature of this dynamical evolution is the fact
that $\lambda_0$ often evolves \emph{non-monotonically},
first rising and then falling again as we pass through the phase transition. 
Indeed, this non-monotonicity in $\lambda_0$ occurs for all 
flow lines which cross the green line in Fig.~\ref{fig:lambda_contour}, 
with values of \il{(\alphabar,\betabar)} lying below and/or to the left of this green line.
As we see from Fig.~\ref{fig:lambda_contour},
such behavior tends to be rather generic across much of the \il{(\alphabar,\betabar)} 
parameter space, 
and is particularly dramatic in cases for which \il{\alphabar <0} and $\betabar$ is relatively small. 

We can understand how this behavior arises as follows.
At early times prior to the phase transition, 
the eigenvalues of our mass matrix are \il{\lambda_0=0} and \il{\lambda_1=M}.
However, as we enter the phase transition, the  
value of $\lambda_0$ inevitably begins to rise, and
when \il{\alphabar<0} the two eigenvalues
$\lambda_{0,1}$ actually begin to approach each other.
There is no level-crossing, however, because 
the mixing between these states ultimately induces a level-repulsion
which kicks $\lambda_1$ toward even higher values
while causing $\lambda_0$ to drop again 
before settling into its final, asymptotic value.
This level-repulsion is particularly strong in cases
for which the two states experience significant mixing 
(\ie, cases in which $\betabar$ is relatively small).
Thus it is ultimately \emph{level repulsion} that lies at the root of the
non-monotonicity of $\lambda_0$.

\begin{figure}[t]
    \begin{center}
    \includegraphics[keepaspectratio, width=0.49\textwidth]{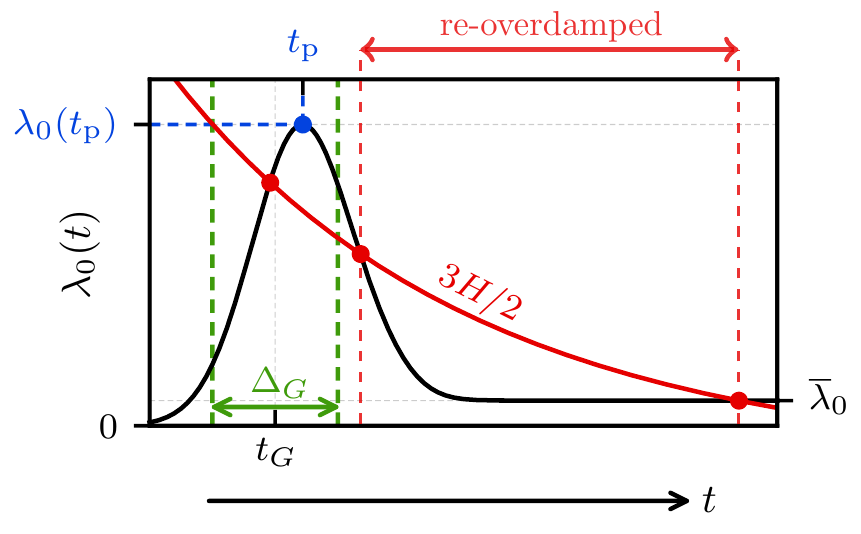}
    \end{center}
    \caption{``Pulse'' behavior for the lightest eigenvalue $\lambda_0$, sketched as a function of time.
    This pulse behavior is ultimately 
    the result of level repulsion between the two scalars in our system, and is particularly severe
    when the two mass eigenvalues approach each other 
    (as occurs when \il{\alphabar<0}) and experience significant mixing 
    (as occurs when $\betabar$ is relatively small). 
    For the special case \il{m_{\rm sum}^2=M^2}, such pulse behavior emerges for all 
     \il{(\alphabar,\betabar)} lying below and/or to the left of 
      the green contour in Fig.~\ref{fig:lambda_contour}.~ 
    As we shall see, this pulse behavior simultaneously gives rise to the two ingredients
    which are critical for building a slingshot for $\phiLN$:
    the phenomenon of re-overdamping and the possibility of a parametric resonance.  }
\label{fig:lambda_resonance}
\end{figure}

This non-monotonic ``pulse'' behavior for $\lambda_0$ is sketched in Fig.~\ref{fig:lambda_resonance}.~
For concreteness, we have also superimposed a Hubble parameter which falls with time as \il{H(t)\sim 1/t}. 
As we shall see, the existence of such a pulse has two important consequences.
Both of these will turn out to be critical ingredients in building our eventual slingshot.

The first of these ingredients is the existence of a ``re-overdamped'' phase in the dynamics of the lighter
scalar field.  Recall that in general, a scalar of constant mass $m$ begins in an 
overdamped phase if \il{3H(t) \geq 2m}
at a sufficiently early time $t$.  During such a phase, the field VEV remains approximately constant,
with very little kinetic energy.  However, since the Hubble parameter falls with time as the
universe evolves, there eventually comes a point at which such a scalar transitions to an underdamped phase
in which \il{3H(t) < 2m}.  The field VEV then begins 
to oscillate around zero.  \emph{However, if the mass of the scalar in question experiences a pulse, 
as shown in Fig.~\ref{fig:lambda_resonance},
it is possible for the scalar to become overdamped once again for a non-trivial interval
of time --- even after having already been underdamped.}
Indeed, we see from Fig.~\ref{fig:lambda_resonance} 
that there are \emph{two} distinct intervals of time during which \il{3H(t) \geq 2 \lambda_0(t)}.
Upon entering this second overdamped phase, the oscillations of the field VEV cease, as expected.
\emph{However, because the field VEV is already oscillating when it enters the re-overdamped phase,
it enters with an initial velocity which is no longer subject to the restoring forces
that would have produced oscillations.}  Such a field VEV thus retains this initial velocity,
subject only to the Hubble friction that eventually brings this velocity to zero.

The existence of a ``pulse'' for the mass $\lambda_0$ 
can also potentially give rise to our second important ingredient:  a parametric resonance that acts during the
initial underdamped phase --- {\it i.e.}\/,  during the pulse itself.
Recall that in general a harmonic oscillator experiences an $n^{\rm th}$-order  
parametric resonance that magnifies the amplitudes 
of successive oscillations if the mass of the oscillator exhibits its own oscillatory behavior whose frequency
is approximately $(2/n)$ times the natural oscillator frequency, with \il{n\in \IZ^+}.
However, this is exactly what can occur here, 
since the pulse described above furnishes us with a changing mass.
Of course, a true parametric resonance requires that the mass experience periodic oscillations,
whereas in our case $\lambda_0$ experiences only a single pulse.  \emph{However, such a pulse mimics 
half an oscillation, and it turns out that even this small segment of an oscillation
is sufficient to induce a parametric resonance provided this pulse has 
an appropriate effective frequency $\omega_{\rm eff}$.}
In general, the effective frequency $\omega_{\rm eff}$ 
of the pulse may be obtained from the curvature of $\lambda_0^2(t)$ 
near \il{t\approx \tp}, where $\tp$  is the time at which the pulse reaches its maximum height,
and is given by~\cite{Dienes:2015bka} 
\beq
     \omega_{\rm eff}^2 ~=~ -4 \left. \frac{\ddot \lambda_0}{\lambda_0}\right|_{t=\tp}~.
\eeq
By contrast, the natural frequency of the oscillating scalar field near \il{t\approx \tp} 
is nothing but $\lambda_0(\tp)$,
since it is the mass of the field that drives the oscillations.  We thus obtain a condition 
for the existence of an $n^{\rm th}$-order parametric resonance~\cite{Dienes:2015bka}: 
\beq
         \left.\left( \frac{\ddot \lambda_0}{\lambda_0^3}\right)\right|_{t=\tp} ~=~ - \frac{1}{n^2}~,~~~~ n\in\IZ^+~.
\label{eq:rescond}
\eeq
Varying the width $\Delta_G$ of the phase transition induces variations in the value of the left side of
this equation. Thus, there exists a discrete set of phase-transition widths $\Delta_G^{(n)}$ for which
this resonance condition is satisfied.
Assuming the width $\Delta_G$ of the phase transition matches one of these 
resonant widths $\Delta_G^{(n)}$,
an $n^{\rm th}$-order parametric resonance therefore enhances the size of the 
field oscillations near the end of the 
underdamped phase that immediately precedes re-overdamping. 

Putting these two ingredients together, we now have the basic recipe 
for our slingshot that propels the lighter mass-eigenstate 
field $\phiLN$ to a super-Planckian VEV.
Choosing an appropriate set of parameters \il{(\alphabar,\betabar)} gives rise to a pulse for the mass $\lambda_0$ of
this field.  If this pulse has an appropriate shape,
as described in Eq.~(\ref{eq:rescond}), the field 
oscillates during the pulse.  Of course, it is critical for the functioning of our 
slingshot that we enter the re-overdamped phase precisely at a moment where the 
oscillating field has its maximum velocity, so that our field
is ``released'' from oscillatory behavior and launched with the maximum possible velocity.
However, it turns out that the condition  
for this to happen is the same as that in Eq.~(\ref{eq:rescond}) which establishes the
parametric resonance in the first place.   Thus, once the parametric resonance is established through
the proper choice of $\Delta_G$,
the resulting pulse has precisely the correct width in order to launch 
the field VEV with maximum velocity into the re-overdamped phase.
Indeed, it turns out that if the width of the phase transition is given by $\Delta_G^{(n)}$, 
appropriate for satisfying the $n^{\rm th}$-order parametric resonance condition,
then our field experiences $n/2$ oscillations within the underdamped phase before 
the eventual ``launch'' into the re-overdamped phase.
As a result, parametric resonances with odd orders $n$ launch their field VEVs in a direction which
is opposite to that for even orders $n$.

\begin{figure}[t!]
    \begin{center}
    \includegraphics[keepaspectratio, width=0.485\textwidth]{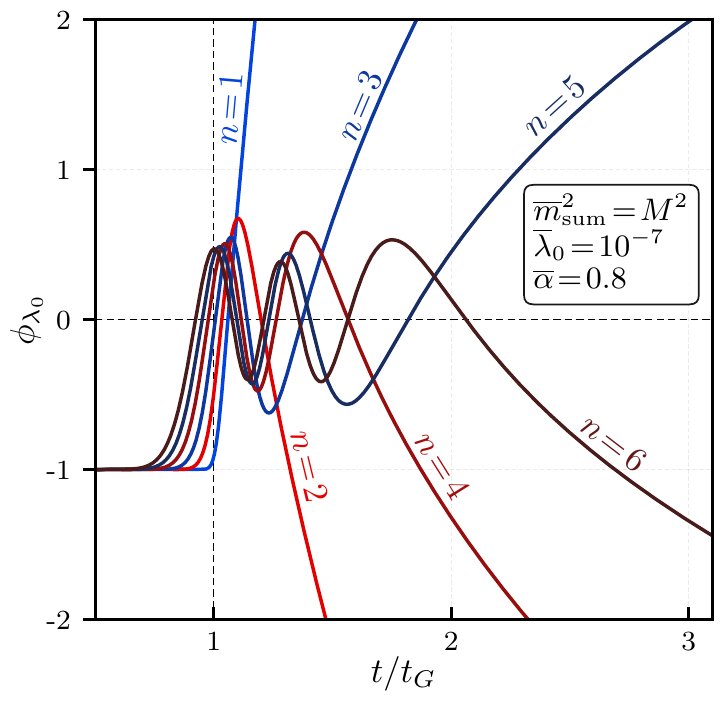}
    \end{center}
    \caption{The time-evolution of the $\phiLN$ field
    in cases where the phase-transition width $\Delta_G$ is equal to
    a resonant width $\Delta_G^{(n)}$, plotted as functions of time 
    for \il{1\leq n\leq 6} and assuming an initial field VEV \il{\phiLN = -M_{\rm P}} for \il{t\ll t_G}.
    (This sign is chosen as a convention so that the primary \il{n=1} resonance
    propels our field VEV toward positive values.)
    For each $n$ we see that our slingshot mechanism successfully propels the field toward a
    large field VEV.~  In this figure (and in all subsequent figures in this paper), all dimensionful
    quantities are to be understood in units of appropriate powers of the Planck scale $M_p$, and we have taken
    \il{M= M_p/3} throughout.  Note that the parameters chosen for this plot --- namely 
    \il{m_{\rm sum}^2 = M^2}, \il{\alphabar= 0.8}, and \il{\lambdabar_0= 10^{-7} M_p} --- correspond 
    to the flow line indicated in red in Fig.~\ref{fig:lambda_contour}.  }
\label{fig:early_fields}
\end{figure}

These points are illustrated in Fig.~\ref{fig:early_fields}, 
which shows our slingshot mechanism in action.  
In this figure, we show the time-evolution of the lighter field $\phiLN$ when the 
phase-transition width $\Delta_G$ is set to $\Delta_G^{(n)}$ for \il{n=1,2,...,6}. 
In each case, our field begins with a VEV at $-M_p$ and 
undergoes \il{n/2} oscillations within the underdamped phase
before being launched toward a field value
whose magnitude vastly exceeds that of its initial value prior to the phase transition.
Indeed, in each case the field is ``launched'' precisely upon entrance  
into the re-overdamped phase. 

It is natural to refer to this mechanism for generating a large field VEV as a ``slingshot.''
In ancient times, prior to the invention of rubber and other elastic materials, a slingshot of the David/Goliath
variety was fashioned by attaching a projectile to a rope, twirling the rope around overhead with
increasing speed, 
and then releasing the projectile at just the right moment so as to launch the projectile in the desired direction.
Our slingshot mechanism is essentially the same:   we too begin with an interval
of periodic oscillations enhanced through a parametric resonance, followed by a ``release'' from the oscillatory
behavior
at just the proper moment so as to propel the projectile forward with maximum velocity.   
Indeed, the projectile in this instance is nothing but the field VEV, and its release from 
oscillations is nothing but the entrance into the re-overdamped phase.
Likewise, the different higher-order resonances correspond to the different points at which the release
can take place, with successive higher orders of resonance alternating between 
forward or backward motion of the projectile.

\begin{figure}[t!]
    \begin{center}
    \includegraphics[keepaspectratio, width=0.485\textwidth]{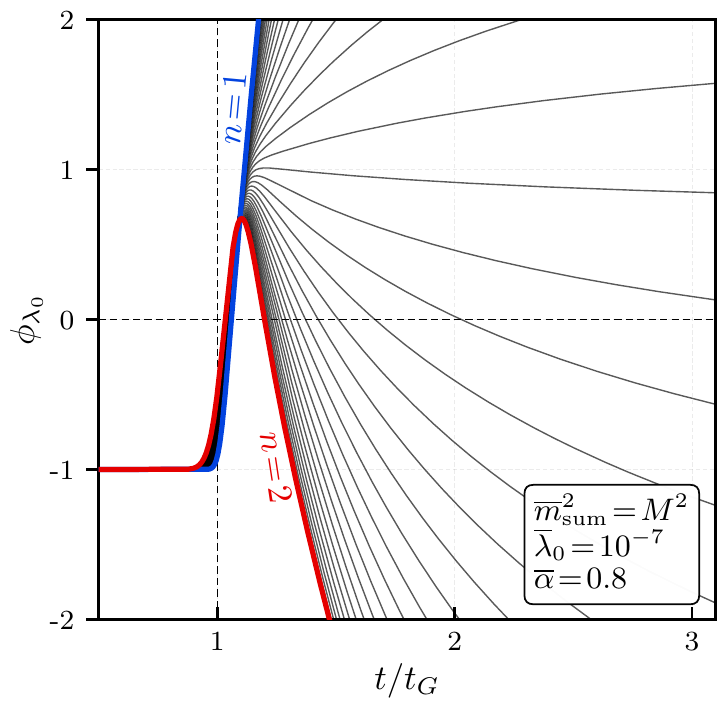}
    \end{center}
    \caption{The \il{n=1} (blue) and \il{n=2} (red) resonance curves from Fig.~\ref{fig:early_fields}, along 
with a spectrum of 48 intermediate curves that describe the behavior of the $\phiLN$ field for 
phase-transition widths $\Delta_G$ lying
at equally spaced intervals between the resonant widths $\Delta_G^{(1)}$ and $\Delta_G^{(2)}$. 
From top to bottom, these curves correspond to phase-transition widths 
\il{\Delta_G = (1-r/49) \Delta_G^{(1)} + (r/49) \Delta_G^{(2)}} with \il{r=1,2,...,48} respectively. 
For reference, the curve whose asymptotic field behavior is almost completely flat corresponds to \il{r=27}. 
We see that even when we are not sitting precisely on a resonant width $\Delta_G^{(n)}$, our slingshot mechanism
is often still capable of propelling our field VEVs to relatively high values.
This ultimately happens because our parametric resonances are quite broad. Thus 
our slingshot mechanism does not require a significant amount of fine-tuning in order
to operate as desired.}
\label{fig:early_fields2}
\end{figure}

At first glance it might seem that our slingshot mechanism is fine-tuned in the sense that we must
be precisely sitting on the parametric resonance, with \il{\Delta_G= \Delta_G^{(n)}}, in order to successfully launch   
our fields to other regions.  However, this is not the case.
In Fig.~\ref{fig:early_fields2}, we show the same \il{n=1} and \il{n=2} resonance curves from Fig.~\ref{fig:early_fields},
along with 48 additional curves that illustrate the behavior of the field $\phiLN$ when we are \emph{not}
sitting on either resonance, but rather are situated \emph{between} these     
resonances.  Indeed, in some sense these curves ``interpolate'' between 
the \il{n=1} and \il{n=2} resonance curves and correspond to phase-transition widths $\Delta_G$ which progress in equal-sized
steps from $\Delta_G^{(1)}$ to $\Delta_G^{(2)}$.
We see that even in cases which are off resonance, our slingshot mechanism continues to 
propel the field VEV toward large values.
In other words, our parametric resonances are quite broad, and our system continues to benefit  
from the existence of these resonances even if it is not finely-tuned to match their parameters.

\begin{figure}[t!]
    \begin{center}
    \includegraphics[keepaspectratio, width=0.485\textwidth]{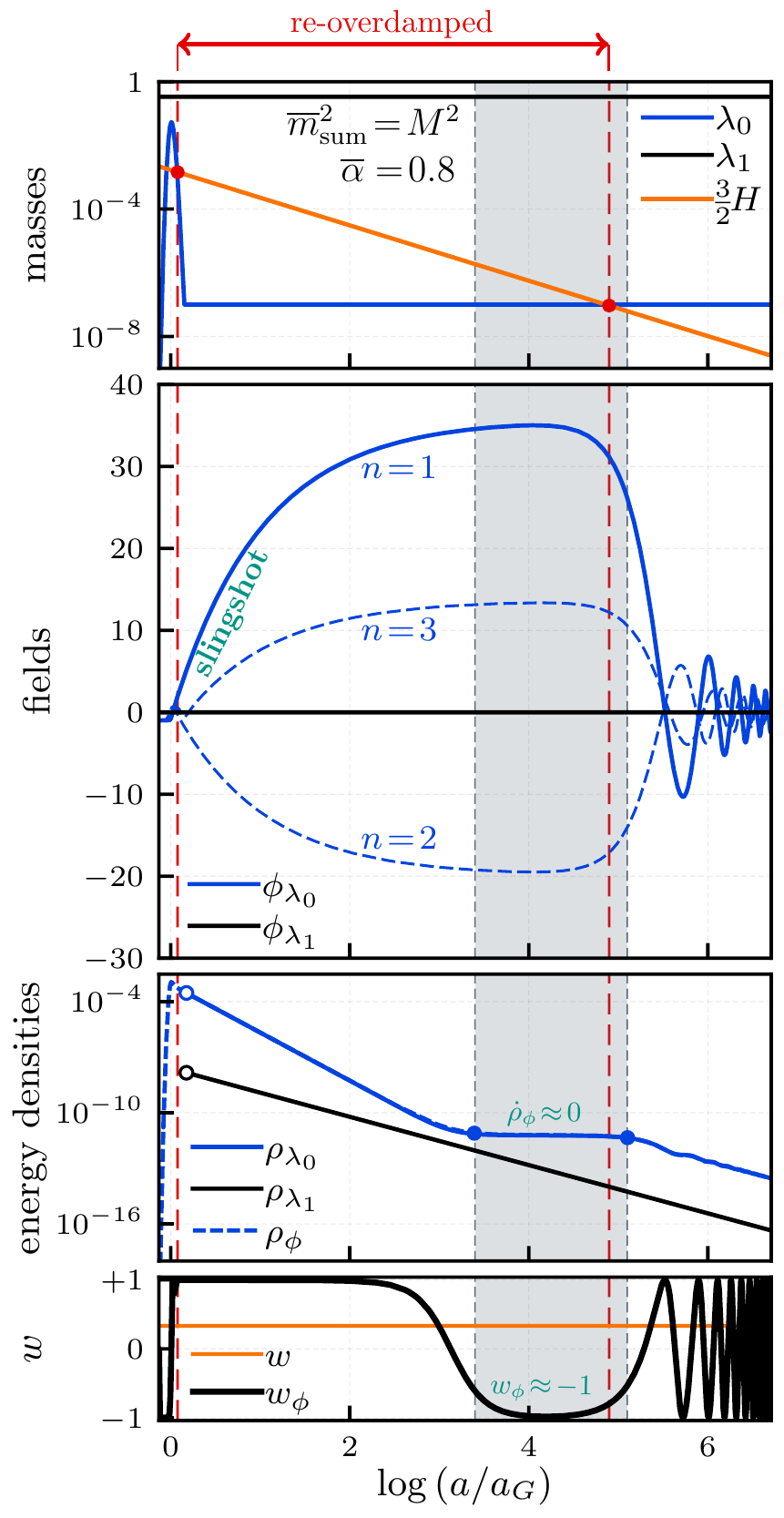}
    \end{center}
    \caption{Dynamical evolution of the scalar sector within
      a background (radiation-dominated) cosmology with a falling Hubble parameter \il{H(t)\approx 1/(2t)}.  
    The different panels from top to bottom indicate the evolution of the
     scalar masses $\lambda_i$, the corresponding fields $\phi_{\lambda_i}$, 
    their energy densities $\rho_{\lambda_i}$, and the equation-of-state parameter $\wphi$
    of the scalar sector.  
    All curves correspond to the case of the \il{n=1} resonance, 
     except the dashed curves which show how the fields would have evolved 
       under the \il{n=2} and \il{n=3} resonances instead.
     For this figure we have taken {\il{M=M_p/3}} and {\il{\lambdabar_0=10^{-7} M_p}} as reference values.
        We observe that {\il{\dot\rho_{\lambda_0}\approx 0}}
         during the time interval shaded in gray, 
     with the equation-of-state parameter for our scalar
       sector remaining near {\il{w_\phi\approx -1}} during this interval.   
}
\label{fig:phirho_subdominant}
\end{figure}

Given this slingshot mechanism, we can now proceed to consider how our scalar sector
would evolve in a cosmological setting, both during and after the slingshot.
In Fig.~\ref{fig:phirho_subdominant} we show the
evolution of the masses $\lambda_i$ (top panel), the fields $\phi_{\lambda_i}$ (second panel), 
and their corresponding energy densities $\rho_{\lambda_i}$ (third panel).
For all panels, the flow of time is indicated in terms of {\il{\log(a/a_G)\equiv \log[a/a(t_G)]}}, the number of 
$e$-folds since the phase transition,
and we have plotted all of these quantities as functions of 
this number under our continuing assumption that our background cosmology is radiation-dominated.
Note that the energy densities plotted in 
Fig.~\ref{fig:phirho_subdominant} are given by 
\beq
 \rhobar_{\lambda_i} ~=~ \half 
     \left(  \dot \phi_{\lambda_i}^2 + \lambdabar_i^2 \phi_{\lambda_i}^2\right)~.
\label{eq:rholambdas}
\eeq
However, these energy densities can be meaningfully associated with the mass-eigenstate
fields $\phiLN$ and $\phi_{\lambda_1}$ only after the phase transition has effectively
concluded and the mixing angle $\theta$ between these states has settled into its
asymptotic, late-time value $\thetabar$.  By contrast, at earlier times, the identities of the mass
eigenstates are continually evolving and changing, and the existence of additional
$\dot\theta$-dependent mixing terms within the total energy density
$\rho_{\phi}$ of the scalar system prevents $\rho_\phi$ from being cleanly separated into
individual contributions of the form appearing in Eq.~(\ref{eq:rholambdas}).
For this reason the separate energy densities $\rho_{\lambda_i}$ are plotted
only after $\dot\theta$ has become sufficiently small and these contributions to $\rho_\phi$
can be meaningfully separated.
The moment at which this occurs is indicated with open circles within this panel, and at earlier times we 
simply plot the total scalar-sector energy density $\rho_\phi$.
Likewise, in Fig.~\ref{fig:phirho_subdominant}
we have also plotted the equation-of-state parameter $\wphi$ of the scalar sector (bottom panel).  
This is defined as
\beq
    w_{\phi} ~\equiv~ \frac{P_{\phi}}{\rho_{\phi}} ~=~ 
    \frac{\half \sum_k \phidot^2_k - 
    \Veff(\phi_i,t)}{\half \sum_k \phidot^2_k + \Veff(\phi_i,t)}~, 
\label{eq:wphi}
\eeq
where $P_{\phi}$ and $\rho_{\phi}$ are the total pressure and energy density
associated with the scalar scalar.  
For all panels of this figure we have chosen 
{\il{M=M_p/3}} as a benchmark value, with {\il{m_{\rm sum}^2=M^2}}, {\il{\alphabar=0.8}}, and 
{\il{\lambdabar_0 = 10^{-7}M_p}}.  
We have also taken our phase transition to have a width 
{\il{\Delta_G=\Delta_G^{(1)}}}.

The results in Fig.~\ref{fig:phirho_subdominant} can be understood as follows.
In the top panel, we see the behavior of the mass eigenvalues $\lambda_{0,1}$ (blue and black curves, respectively), 
along with 
the Hubble curve superimposed (red).  The Hubble parameter scales as $1/(2t)$, as expected for
a radiation-dominated cosmology, while $\lambda_1$ remains close to $M_p$
across the full time interval shown in the figure.
However $\lambda_0$ clearly exhibits the ``pulse'' we have discussed, 
along with a relatively long subsequent period of re-overdamping.
This re-overdamping period is demarcated 
in Fig.~\ref{fig:phirho_subdominant} 
as that period existing between the 
vertical dashed red lines.

In the second panel, we show the corresponding 
behavior of the mass-eigenstate fields $\phi_{\lambda_{0,1}}$.
As discussed above,
the heavier field $\phi_{\lambda_1}$ (solid black curve) remains underdamped and thus
experiences damped oscillations throughout
the time interval shown, but the amplitudes
of these oscillations are always sufficiently small that these
oscillations are not readily evident in this figure. 
By contrast, the VEV of the lighter
field $\phi_{\lambda_0}$ (solid blue curve) experiences a dramatic change as a result of the
``slingshot'' dynamics:  this field is endowed with a huge velocity upon emerging from the
phase transition and for this choice of model parameters is ultimately propelled to a value around
\il{\phi_{\lambda_0}\approx 35M_p} after only a few \e-folds.
Indeed, it is only after the period of re-overdamping ends that this field begins to exhibit
damped oscillations, as expected.  

For purposes of comparison, we also display the
corresponding $\phi_{\lambda_0}$ curves (dashed blue curves) that would result if we had taken 
{\il{\Delta_G=\Delta_G^{(2)}}} or {\il{\Delta_G=\Delta_G^{(3)}}} rather than {\il{\Delta_G=\Delta_G^{(1)}}}.
We see that for each of these higher-order resonances, our slingshot mechanism likewise 
endows $\phi_{\lambda_0}$ with a large, trans-Planckian field VEV --- albeit a VEV which is
somewhat smaller than that obtained for the primary resonance.

In the third panel of Fig.~\ref{fig:phirho_subdominant}, we show the evolution of the energy
densities $\rho_{\lambda_{0,1}}$.  The energy density
$\rho_{\lambda_1}$ associated with the heavier scalar (black curve) declines steadily
as \il{\rho_{\lambda_1}\propto a^{-3}} across the
entire time interval shown in the figure, as appropriate for a field which behaves
like massive matter.  
By contrast, the energy density $\rho_{\lambda_0}$ of the lighter
field (blue curve) evolves in non-trivial way.  
Immediately upon emerging from the phase transition, the energy density of this field 
is almost entirely kinetic.  It thus redshifts much more rapidly, 
dropping as \il{\rho_{\lambda_0}\propto a^{-6}}.
However, 
as the lighter field VEV
approaches its apex, 
the dissipation 
of the corresponding energy density ceases 
almost entirely.
Indeed, at this point the velocity of the lighter field approaches zero,
whereupon our field is fully re-overdamped with no residual velocity remaining from the slingshot.
This is thus an epoch in which 
the energy density of the lighter field behaves effectively as vacuum energy,
with {\il{\dot\rho_{\lambda_0}\approx 0}}. 
Because of the importance of this epoch in our eventual discussion, we have
shaded this 
{\il{\dot\rho_{\lambda_0}\approx 0}}
epoch
with a gray background in Fig.~\ref{fig:phirho_subdominant}.~
This epoch only ends when the period of re-overdamping ends, whereupon this 
field begins to oscillate again and $\rho_{\lambda_0}$ begins to fall accordingly as $1/a^3$.

In the bottom panel of Fig.~\ref{fig:phirho_subdominant} we plot two quantities:  the
effective equation-of-state  parameter $w_\phi$ for our scalar system (black curve), and the 
equation-of-state parameter $w$ for this entire example universe (orange). 
The behavior of $w_\phi$
follows directly from the properties we have seen in the previous panels.
Because {\il{\rho_{\lambda_0}\gg \rho_{\lambda_1}}}
at all times after the phase transition, $w_\phi$ is essentially determined by the
behavior of $\phiLN$.  At early times --- {\it i.e.}\/, for {\il{a \ll a_G}} --- we have {\il{w_\phi \approx -1}}.
However, the scalar sector becomes kinetic-energy dominated, with {\il{w_\phi \approx +1}},
immediately after $\phiLN$ is released from the slingshot.  As we enter the fully re-overdamped
region shaded in gray, $w_\phi$ drops to $-1$, signifying the passage to a vacuum-energy
dominated phase for the scalar sector.  When the period of re-overdamping eventually ends and
$\phiLN$ begins to behave like massive matter, $w_\phi$ begins to oscillate 
around {\il{w_\phi\approx 0}}.
Of course, despite this behavior for $w_\phi$, we see that 
the equation-of-state parameter $w$ for the universe as a whole remains fixed at {\il{w = +1/3}}.
This reflects our original assumption that 
we are operating within a background cosmology which is radiation-dominated, as consistent with the falling
Hubble curve indicated in the top panel of Fig.~\ref{fig:phirho_subdominant}.

\section{Building an inflationary slingshot cosmology\label{sec:Inflation}}


In this section, we discuss how the scalar sector and slingshot mechanism 
described in Sect.~\ref{sec:TheScalarSector} can
become the core elements of an inflationary cosmology.
In Sect.~\ref{sec:subsect1}, we describe how a full slingshot cosmology
can be constructed and we demonstrate that such a cosmology can indeed give rise
to an inflationary epoch.
Then, in Sect.~\ref{sec:subsect2}, we show that such an inflationary epoch can 
can have a duration sufficient to address the
horizon and flatness problems.

\subsection{From scalar sector to inflationary slingshot cosmology \label{sec:subsect1}}


In Sect.~\ref{sec:TheScalarSector}, we 
considered a system of two scalar fields which undergo
a cosmological phase transition
and demonstrated a mechanism in which a
parametric resonance followed by a re-overdamping phase
together conspire to ``slingshot'' the VEV of the lighter field 
to super-Planckian values.
We even found that such a system can give rise to many of the features
normally associated with cosmological 
inflation, with the lighter scalar field effectively experiencing something akin to 
``slow-roll'' behavior, with 
{\il{\dot \rho_{\lambda_0}\approx 0}}.
Indeed, this situation arises within the gray shaded region
in Fig.~\ref{fig:phirho_subdominant}.~
We even demonstrated that the energy associated with our scalar sector as a whole 
behaves as vacuum energy during this period, with a corresponding
equation-of-state parameter {\il{w_\phi\approx -1}}. 

Unfortunately, this period of evolution is ultimately not inflationary.
Inflation would require a slowly varying Hubble parameter {\il{|\dot H| \ll H^2}},
but instead our Hubble parameter is falling with \il{H(t)\approx 1/(2t)}, as consistent with our 
original assumption of a fixed radiation-dominated background cosmology.
Likewise, we have seen that the equation of state for the entire universe is correspondingly 
fixed at {\il{w= +1/3}}, as expected under radiation-domination.
Of course, in Sect.~\ref{sec:TheScalarSector} 
we fixed the background cosmology in this way so that we could focus on the 
scalar sector.  However, by embedding our scalar sector within
a fixed background cosmology, we were effectively disregarding the 
gravitational \emph{backreaction} of this cosmology on the scalar field dynamics.
In other words, we were implicitly assuming that the scalar sector 
is only a \emph{subdominant}\/ component of our total cosmology, with a 
total scalar-sector energy density which is much less than the total energy density of the universe.

In order for our slingshot effect to potentially trigger an inflationary 
phase for the universe, we must instead enlarge our perspective 
by treating the scalar sector as the \emph{dominant}\/ component.
However, this change in perspective immediately introduces several complications
that must be considered.

First, we must take care to track the total energy flow within 
our cosmological setup.  Recall that our scalar sector begins without energy, 
and only gains energy through the cosmological phase transition.   
Indeed, it is this phase transition which is ultimately 
responsible for the explicit time-dependence in the effective potential $V_{\rm eff}(\phi_i,t)$ 
in Eq.~(\ref{eq:Veff}).
However, if this energy is now to be considered the dominant 
component of energy density in the universe, then we must account 
for its \emph{source}.
In particular, since the energy density $\rho_\phi$ of our scalar sector has a time-evolution
governed by
\beq
\dot{\rho}_{\phi} + 3H\left(1 + w_{\phi}\right)\rho_{\phi} ~=~ +\frac{\p \Veff}{\p t} ~,
\eeq
there must also be a generic \emph{source sector} with energy density $\rhoS$ governed by
\beq\label{eq:sourceeq}
\dot{\rho}_{S} + 3H\left(1+w_S\right)\rhoS ~=~ -\frac{\p \Veff}{\p t} ~.
\eeq
This will therefore balance the total energy budget in the \il{H\rightarrow 0} limit (or equivalently
in the co-moving frame), as required.  It is of course natural to imagine that \il{w_S=-1}, so 
that this sector consists  
of vacuum energy, and we shall make this choice throughout the rest of this paper.
Indeed, we shall further assume that $\rhoS$ 
is negligible after the phase transition, and thus its details are irrelevant for the physics
after this transition.

In a similar vein, we may also wish to have a radiation sector of some sort 
to trigger the phase transition as the universe cools.
Thus, we shall henceforth consider a minimal 
cosmological model consisting of three sectors:
the scalar sector as outlined in Sect.~\ref{sec:TheScalarSector}, a source 
sector as described above,
and a radiation sector with total energy density $\rhoR$. 
Indeed, our discussion in Sect.~\ref{sec:TheScalarSector} implicitly assumed the relation
\beq
    \rhoS + \rho_\phi ~\ll ~ \rhoR ~,
\label{eq:assum1}
\eeq
but we now wish to consider the more general situation in which
\beq
    \rhoS + \rho_\phi ~\gsim ~ \rhoR ~
\label{eq:assum2}
\eeq
during and after the phase transition, when our slingshot operates.
In fact, the specific initial conditions in 
Sect.~\ref{sec:TheScalarSector}
imply that {\il{\rho_\phi=0}} prior to the phase transition, whereupon
the assumption in Eq.~(\ref{eq:assum2}) reduces to the assumption that \il{\rhoS \gsim \rhoR}.

This in turn leads to our second complication. 
In Sect.~\ref{sec:TheScalarSector}, our analysis of the slingshot assumed 
a Hubble parameter which falls as \il{H(t)\approx 1/(2t)}, regardless
of the dynamics of the scalar sector.
However, if we now wish to incorporate the situation in Eq.~(\ref{eq:assum2}),
then the scalar and source sectors all contribute non-negligibly, and in general
$H(t)$ is given by the Friedmann equation
\beq\label{eq:Hdominant}
  H ~=~ \sqrt{\frac{\rho_{\phi} + \rhoS + \rhoR}{3M_p^2}} ~.
\eeq
This then introduces a gravitational back-reaction on the dynamics of scalar system,
so that its equations of motion are rendered non-linear. 
As a result, we expect the dynamics discussed in Sect.~\ref{sec:TheScalarSector} to be significantly altered.
In particular, we now must determine the extent to which the critical ingredients of our slingshot mechanism
in Sect.~\ref{sec:TheScalarSector} --- \eg, the re-overdamping phase, the parametric resonance, {\it etc}\/.  
--- survive 
this gravitational backreaction.  After all, it is \emph{a priori} possible that our 
slingshot mechanism as a whole fails to survive this change.  
Moreover, even if these features do survive, 
it is possible that our parametric resonances might become much more narrow and thereby require
an extreme fine-tuning in the value of $\Delta_G$.  
This too would be an undesirable outcome.

To investigate these possibilities, we repeat the calculations leading to 
Fig.~\ref{fig:phirho_subdominant}, now under the assumption given in Eq.~(\ref{eq:assum2}).
In particular, we shall consider two particular cases of Eq.~(\ref{eq:assum2}), one in which
{\il{\rho_R(t_\medwhitestar) = M_p^4}} and the second in which
{\il{\rho_R(t_\medwhitestar)= 10^{-3} M_p^4}}.
We shall also continue to assume
the same parameter choices as for Fig.~\ref{fig:phirho_subdominant}, namely
{\il{M=M_p/3}}, {\il{\mbar_{\rm sum}^2=M^2}}, {\il{\alphabar=0.8}}, and {\il{\lambdabar_0=10^{-7}M_p}}.

\begin{figure}[t]
    \begin{center}
    \includegraphics[keepaspectratio, width=0.49\textwidth]{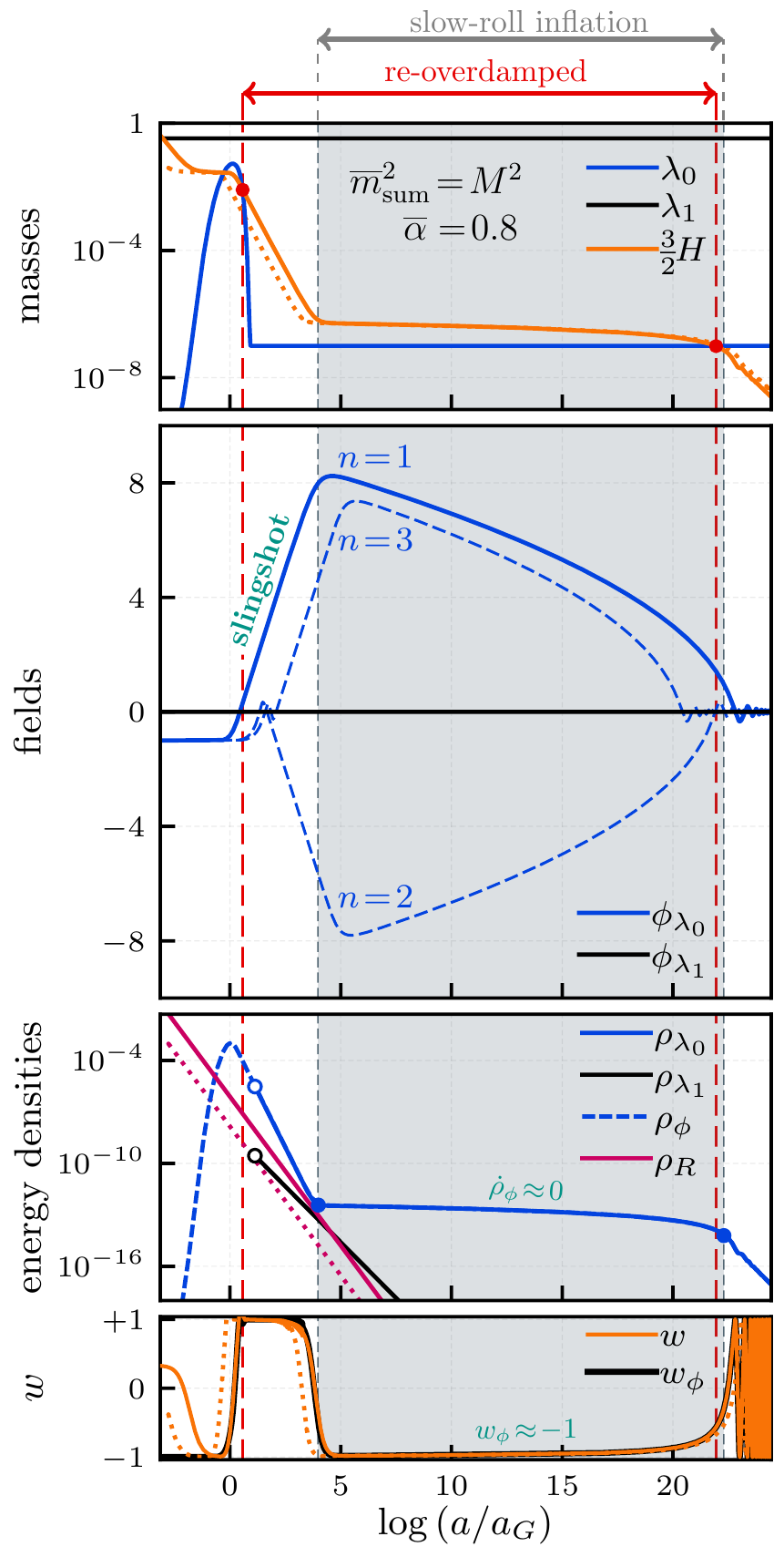}
    \end{center}
    \caption{
    Same as Fig.~\ref{fig:phirho_subdominant}, except within the context 
    of a cosmology in which the energy density of the scalar sector dominates.
    We see that the resulting gravitational backreaction not only
    preserves our slingshot 
    but also leads to a period of slow-roll inflation (shaded in gray).}
    \label{fig:phirho_dominant}
\end{figure}

Our results are shown in Fig.~\ref{fig:phirho_dominant}, where
the solid (dotted) curves for $H(t)$, $\rho_R(t)$, and $w(t)$ correspond
to the larger (smaller) value of $\rho_R(t_\medwhitestar)$ listed above. 
Several features immediately become apparent upon comparing these
results with those in Fig.~\ref{fig:phirho_subdominant}.~ 
First, we note from the top panel that the Hubble curve still crosses the lighter
mass $\lambda_0$ at multiple points, giving rise to several damping transitions.
In other words, the re-overdamping phase appears to survive gravitational backreaction.  
Second, examining the field evolution, we find that the parametric resonance also survives,
including the higher-order resonances.  
Thus, our slingshot remains intact:  we continue to have a relatively
long period of re-overdamping at the beginning of which the VEV of the lighter field $\phi_{\lambda_0}$
is propelled to large, trans-Planckian values.
Indeed, we see that the period of re-overdamping is now significantly extended beyond what it was
in Fig.~\ref{fig:phirho_subdominant}, lasting for more than three times as many $e$-folds.

As before, the magnitude of the field VEV $\phi_{\lambda_0}$ continues to grow, 
in this case ultimately reaching a maximum 
value {\il{\phi_{\lambda_0}\approx 8 M_p}},
until the kinetic energy of the field is depleted by Hubble friction. 
This once again leads to an epoch during which
{\il{\dot\rho_{\lambda_0}\approx 0}}, as shaded in gray.   
Indeed, during this epoch the lighter field satisfies the slow-roll conditions.
Moreover, the energy density in the lighter scalar field continues 
to dominate that of the heavier scalar field, with {\il{\rho_{\lambda_0}\gg \rho_{\lambda_1}}} throughout this interval.
We therefore once again have {\il{w_\phi\approx -1}}.

However, in stark contrast to what occurs in Fig.~\ref{fig:phirho_subdominant}, the total energy density for the
universe is now dominated by the scalar field dynamics.  {\it Thus the
transition to a vacuum-dominated 
scalar sector now indicates the beginning of a truly inflationary epoch,
with the lighter scalar field $\phi_{\lambda_0}$ serving as the inflaton.}
In other words,
the epoch shaded in gray in Fig.~\ref{fig:phirho_dominant} experiences {\it slow-roll inflation}\/,
with {\il{w\approx w_\phi\approx -1}} throughout this epoch.
This is consistent with the fact that the Hubble parameter in the top panel of Fig.~\ref{fig:phirho_dominant}
is now approximately constant during this period --- the same fact 
which is responsible for lengthening the 
period of re-overdamping relative to what it was in Fig.~\ref{fig:phirho_subdominant}, and thereby 
extending our period of inflation to a larger number of $e$-folds than would otherwise have occurred.

We thus conclude that 
\emph{an inflationary epoch can emerge from 
field configurations which, at first glance, do not lead to inflation}.  
Indeed, it is the detailed properties of the cosmological phase transition 
which create the slingshot that propels the VEV of the inflaton field $\phiLN$ to the 
super-Planckian values from which inflation then emerges.

The gravitational backreaction has additional important effects. 
For example, during the slingshot launch of the lighter scalar field $\phi_{\lambda_0}$,
as the VEV of this field is growing toward its maximum value, 
the kinetic-energy density of this field dominates that of the scalar sector
and thus dominates the Hubble damping.
We thus find that \il{H\approx |\phidot_{\lambda_0}|/(\sqrt{6}M_p)}, whereupon
we see that $\phiLN$ evolves linearly with respect to the 
number of \e-folds \il{\Nkin \equiv \int H dt} during this period, \ie, 
\beq
\frac{d\phiLN}{d\Nkin} ~\approx~ \zeta \sqrt{6}M_p ~,
\label{eq:dphidNkin}
\eeq
with an overall sign {\il{\zeta\equiv (-1)^n \,{\rm sgn}({\cal A}_{\phi})}} which depends on 
the order $n$ of the resonance and the initial value ${\cal A}_\phi$ of $\phi_0$, 
as defined below Eq.~(\ref{eq:eqnsofmotion}).
In this sense, the ``velocity'' $d\phiLN/d\Nkin$
of the field as it is launched is independent of
the particular resonance involved. 
Likewise, we see that the maximum field VEV to which our lighter scalar field is launched
is smaller in Fig.~\ref{fig:phirho_dominant} than it was in
Fig.~\ref{fig:phirho_subdominant}.~ 
This loss of efficiency is not surprising, since the damping \il{H\sim \sqrt{\rho_{\phi}}}
is now directly determined by the amount of energy density in the scalar sector.

Consulting Fig.~\ref{fig:phirho_dominant}, we can also see the effect of varying $\rho_R(t_\medwhitestar)$.
As long as \il{\rho_R(t_\medwhitestar) \lsim M_p^4}, it turns out that $\rho_R \lsim \rho_{\lambda_0}$ at the time
when inflation begins.   Thus the radiation component of the total energy density is already subdominant by this point, and changing $\rho_R$ will have essentially no effect on the resulting inflationary epoch. 
Of course, if $\rho_R$ were to {\it exceed}\/ $\rho_{\lambda_0}$ at this time,
then the onset of inflation would be delayed until the later time at which $\rho_R$ finally falls below 
$\rho_{\lambda_0}$.   However, such a situation cannot happen because this greater value of 
$\rho_R$ would imply a super-Planckian radiation energy density prior to the phase transition, which is of course unphysical.
The solid pink line for $\rho_R$ shown in Fig.~\ref{fig:phirho_dominant} is therefore the maximum 
value for $\rho_R$ that can be adopted in any self-consistent cosmology.
On the other hand, our choice for $\rho_R(t_\medwhitestar)$ does have an effect on the physics {\it prior}\/ to
the phase transition, most notably as it concerns the overall equation-of-state parameter $w(t)$. 
However, all of these effects are transient, 
existing only for a short period of time until the phase transition sets in.
We thus conclude that  
the physics of our inflationary epoch
is largely independent of the particular sub-Planckian choice for $\rho_R(t_\medwhitestar)$,
and 
we shall therefore adopt the simplifying assumption that {\il{\rho_R\ll \rho_S+\rho_\phi}} in what follows.

\begin{figure}[b!]
    \begin{center}
    \includegraphics[keepaspectratio, width=0.48\textwidth]{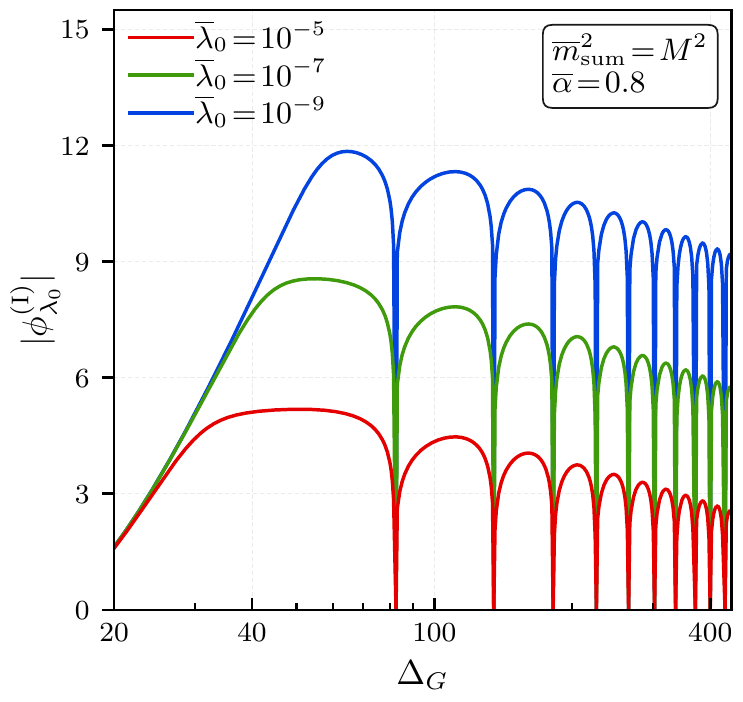}
    \end{center}
    \caption{
        The maximum field VEV $|\phiLNI|$ ultimately reached after the slingshot, 
        plotted as a function of the phase-transition width $\Delta_G$ for several different late-time 
        eigenmasses $\lambdabar_0$.  From left to right, the different peaks within each curve correspond to the
        successive higher-order parametric resonances driven by 
        the mass-generating phase transition.  We see that each of these peaks (particularly 
        that associated with the left-most primary \il{n=1} resonance) is quite broad, with essentially the same large VEV 
        realized for a relatively large range of phase-transition widths $\Delta_G$ near their resonant values. 
        Thus very little fine-tuning is required 
        in order for our slingshot mechanism to capture the benefits of these parametric 
        resonances.}
    \label{fig:dGres}
\end{figure}

We have not yet investigated the extent to which gravitational backreaction
affects one important remaining feature of our slingshot mechanism.  Recall that in the discussion
surrounding Fig.~\ref{fig:early_fields2}, we observed that the slingshot is
not especially fine-tuned in $\Delta_G$.
However, it is possible --- as with the other phenomena presented 
in Sect.~\ref{sec:TheScalarSector} ---  that this property is spoiled
by the gravitational backreaction.  
If so, this would be problematic, as such a fine-tuning would make it far less likely
for realistic slingshot models to exhibit appropriate values of $\Delta_G$.
However, this is ultimately not the case. 
In Fig.~\ref{fig:dGres} we plot the maximum inflaton field VEV $|\phiLNI|$
that is ultimately reached after the slingshot as a function of the phase-transition
width $\Delta_G$  for several different late-time eigenmasses $\lambdabar_0$.
As we see from Fig.~\ref{fig:dGres}, the parametric resonances 
generally continue to be extremely broad,
a feature which remains valid regardless of the mass $\lambdabar_0$ of the inflaton.
In fact, each of the resonances in Fig.~\ref{fig:dGres} is so broad that
a fine-tuning is necessary in order to {\it avoid}\/ resonant behavior.
Our slingshot mechanism can therefore operate with a wide variety of
phase-transition widths $\Delta_G$, and is therefore not particularly sensitive to 
this aspect of the phase transition.

\FloatBarrier
\subsection{Shooting further\label{sec:subsect2}}


Thus far we have demonstrated that our slingshot mechanism --- properly embedded within
an appropriate cosmological framework --- is capable of
giving rise to an inflationary epoch.
Indeed, given the parameters 
underlying the plots in Fig.~\ref{fig:phirho_dominant},
we have seen that the resulting inflationary epoch lasts for approximately
20~$e$-folds.
However, we have yet to explore the full 
parameter space of our model and thereby assess how many \e-folds
of inflation may ultimately be obtained through this mechanism.
In particular, we seek to know whether we can exploit our slingshot mechanism in order to reach
the target {\il{N_{\rm inf}\sim {\cal O}(50-60)}} needed to address the horizon and flatness problems. 

\begin{figure*}[t]
    \begin{center}
        \includegraphics[keepaspectratio, width=\textwidth]{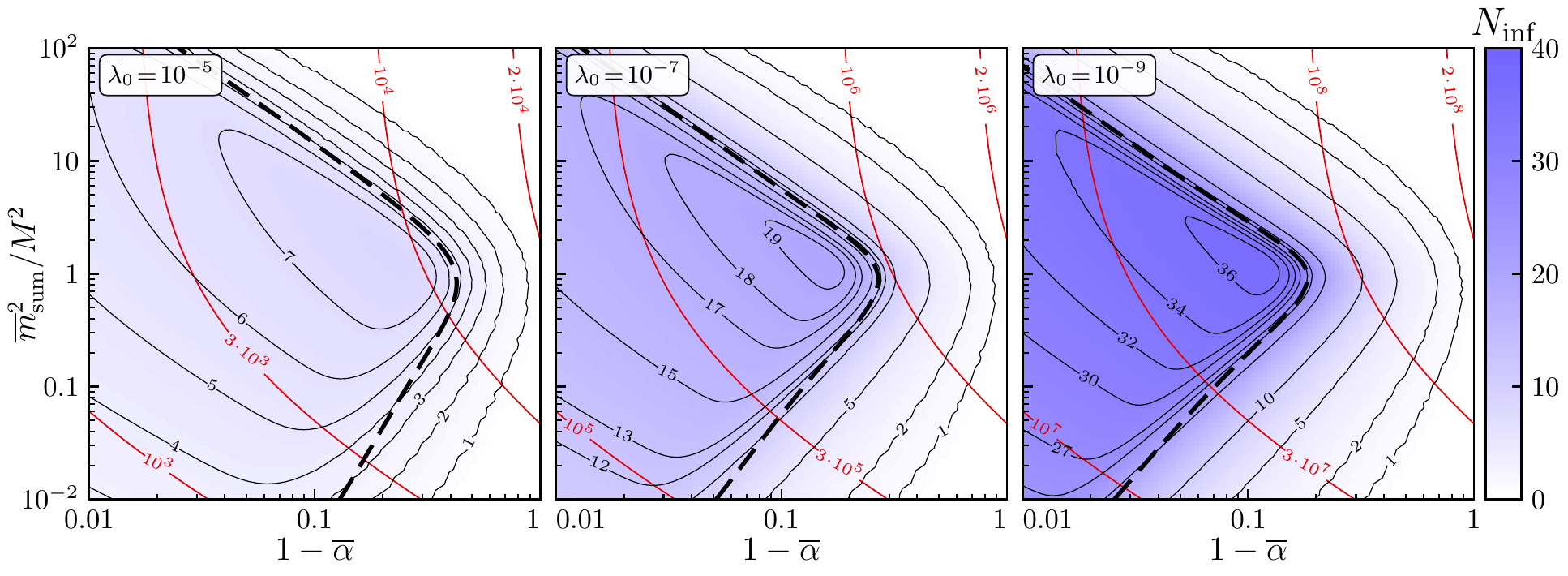}
    \end{center}
    \caption{
        Contours (black curves) showing the number of \e-folds of inflation $\Ninf$ 
        produced by our slingshot, 
        starting from {\il{\phi_{\lambda_0}={\cal A}_\phi = -M_p}} and {\il{\dot\phi_{\lambda_0}=0}}, 
        plotted within the $(\alphabar,\mbar_{\rm sum}^2)$ plane assuming the primary ({\il{n=1}}) resonance.
        The different panels correspond to different choices of the inflaton mass $\lambdabar_0$.  
        In each panel, the thick black dashed curve separates the regions in which
        the total energy density immediately prior to inflation 
           is dominated by that of the lighter field (left) versus the heavier field (right). 
        Contours of the mass quotient 
        \il{\Q\equiv \lambda_0(\tp)/\lambdabar_0} are also shown (red).
        In each case, we see that $\Ninf$ is maximized immediately to the left 
         of the thick black dashed curve, near
        {\il{\alphabar\approx 0.9}} and {\il{\mbar_{\rm sum}^2 \approx M^2}}, 
        and that this maximum value of $\Ninf$ increases with decreasing inflaton mass $\lambdabar_0$ (or 
        increasing ${\cal Q}$). 
    }
    \label{fig:Ninf}
\end{figure*}

Towards this end, we shall now perform a more systematic exploration of the parameter 
space of our slingshot model in order to identify the regions within which $\Ninf$ is maximized.    
Our results are shown in Fig.~\ref{fig:Ninf}.~
In this figure we display contours of the number $\Ninf$
of $e$-folds of inflation 
produced by our slingshot, 
starting from \il{\phi_{\lambda_0}={\cal A}_\phi = -M_p} and \il{\dot\phi_{\lambda_0}=0} prior to the 
phase transition.
These results are obtained by numerically solving 
Eqs.~\eqref{eq:eqnsofmotion} and~\eqref{eq:sourceeq} at each point in parameter space, with $H(t)$ given by 
Eq.~\eqref{eq:Hdominant}.  
From left to right, the results shown in the three panels of the figure respectively correspond 
to the late-time inflaton-mass values {\il{\lambdabar_0/M_p = \lbrace 10^{-5}, 10^{-7}, 10^{-9}\rbrace}}, 
and   we have taken the width of the phase transition to
be {\il{\Delta_G = \Delta_G^{(1)}}} in each case.

Given the results in Fig.~\ref{fig:Ninf}, 
we see that for each choice of $\lambdabar_0$ there actually exists a global maximum for $\Ninf$ 
within the $(\alphabar, \mbar_{\rm sum}^2)$-space.
The location of this maximum is largely insensitive to the value of $\lambdabar_0$, and occurs
around \il{\mbar_{\rm sum}^2\approx M^2} and \il{\alphabar\approx 0.9} in 
each panel of this figure.  
We shall therefore adopt these as benchmark values for all future analyses in these paper.
We also see that the global maximum value of $\Ninf$ actually increases with decreasing $\lambdabar_0$
--- a fact already suggested in Fig.~\ref{fig:dGres} on the basis of the corresponding values 
of $|\phiLNI|$ reached in each case.

The presence of such a global maximum for $\Ninf$ stems from 
the interplay between two competing considerations.  
The first of these considerations is the value of the peak mass $\lambda_0(\tp)$ at top of the pulse, 
as this determines the velocity with which the field is released from the slingshot.  In order
to indicate how $\lambda_0(\tp)$ varies as a function of $\mbar_{\rm sum}^2/M^2$ and 
$\alphabar$ for fixed $\lambdabar_0$, we have also included within each panel of Fig.~\ref{fig:Ninf}
contours (red curves) of the 
dimensionless ratio   
\beq
  \Q ~\equiv~ \frac{\lambda_0(\tp)}{\lambdabar_0}~.
  \label{eq:Qdef}
\eeq
Note that since $\lambdabar_0$ is fixed within 
each panel, contours of $\Q$ are also essentially contours of $\lambda_0(\tp)$.
In general, we observe that within each panel, both $\Q$ and $\lambda_0(\tp)$ increase with 
$\mbar_{\rm sum}^2/M^2$ and with $1-\alphabar$ throughout the parameter space shown. 

The second consideration   which has a significant impact on the value of $\Ninf$ 
is the energy density  
$\rho_{\lambda_1}$ of the heavier field. 
In the regions of parameter space within which $\rho_{\lambda_1}$ represents a significant 
fraction of $\rho_\phi$ at the onset of inflation, 
$\Ninf$ is significantly suppressed.  
In order to illustrate the impact of this suppression, we have included a 
black dashed curve 
within each panel of Fig.~\ref{fig:Ninf} which separates the
regions (to the left and right, respectively) in which 
either $\rho_{\lambda_0}$ or $\rho_{\lambda_1}$ dominates the other at the onset of inflation.
We immediately see that 
$\Ninf$ is significantly suppressed when $\rho_{\lambda_1}$ dominates.  
Thus, within each panel of 
the figure, we see that $\Ninf$ increases with $\lambda_0(\tp)$ (\ie, with
$1-\alphabar$ and $\mbar_{\rm sum}^2/M^2$) ---  but only up 
to the point at which we cross this black dashed curve and this suppression sets in. 
It is for this reason that each value of $\lambdabar_0$ or ${\cal Q}$ leads 
to a global maximum for $N_{\rm inf}$.

Note that requiring that $\rho_{\lambda_1}\lsim \rho_{\lambda_0}$ at the onset
of inflation also places an upper bound on the  
{\it initial}\/ energy density $\rho_{\lambda_1}$ (or equivalently $\rho_1$)
at times $t_\medwhitestar$ prior to the phase transition.
To determine this bound analytically, we begin by noting that
the mass of this field is sufficiently large that this field is always highly
underdamped. Thus \il{\rho_{\lambda_1}\propto a^{-3}} at all times other than during
the phase transition. 
Moreover, relative to the timescales associated with the rapidly oscillating heavy field,
the phase transition is essentially adiabatic and thus does not significantly perturb the 
time-evolution of this field away from this scaling behavior [assuming, of course, that the contribution
to $\rho_{\lambda_1}$ generated by the phase 
transition is less than the initial $\rho_{\lambda_1}(t_\medwhitestar)$].
In order to leave our inflationary epoch undisturbed, we therefore require that
\beq\label{eq:phi1dotsubdominance}
\il{\rho_{\lambda_1}(t_\medwhitestar)\,    e^{-3N} ~\lesssim~ \frac{1}{2}\lambdabar_0^2[\phiLNI]^2} ~ ,
\eeq
where $N$ is the number of \e-folds
between $t_{\medwhitestar}$ and the beginning of inflation.  
We can estimate $N$ by noting that the slingshot occurs 
approximately at the time \il{t_G + \Delta_G}, and therefore
\beq
N ~\approx~ \frac{2}{3(1+\langle w\rangle_{\rm pre})}\log\left(\frac{t_G + \Delta_G}{t_{\medwhitestar}}\right) + \Nkin \ ,
\eeq
where $\langle w\rangle_{\rm pre}$ represents a rough 
average value of the equation-of-state parameter $w(t)$ 
prior to the phase transition 
 and where $\Nkin$ represents the number of $e$-folds during the kination phase prior to inflation.
Using an approximation for $N_{\rm kin}$ to be derived in Eq.~\eqref{eq:Nkinreduction},
we then find
that Eq.~\eqref{eq:phi1dotsubdominance} places a bound on the initial energy density:
\beq
\rho_{\lambda_1}(t_{\medwhitestar}) ~\lesssim~ \frac{1}{2}\,\lambdabar_0^2\,
    \left(\frac{t_G+\Delta_G}{t_{\medwhitestar}}\right)^{2/(1+\langle w\rangle_{\rm pre})}\! \Q \,|\Aphi\phiLNI|  \ .
\label{rho1bound}
\eeq
Moreover, while a precise value of 
$\langle w\rangle_{\rm pre}$ 
depends on model-specific details concerning our initial source and radiation sectors,
we can obtain a conservative estimate 
by taking $\langle w\rangle_{\rm pre}$ equal to its maximum value, which in this case
is $+1/3$.
Of course, the initial conditions we have chosen in Sect.~\ref{sec:TheScalarSector} 
imply that {\il{\rho_{\lambda_1}(t_\medwhitestar)=0}}, so the bound in Eq.~(\ref{rho1bound}) is always satisfied.

Overall, however, 
we see from Fig.~\ref{fig:Ninf} 
that our slingshot mechanism leads to
an inflationary epoch spanning a significant number of \e-folds 
within sizable regions of the parameter space shown. 
Moreover, comparing the results across the three panels of this figure, we also observe that 
the maximum value of $\Ninf$ is largest when $\lambdabar_0$ is small and the corresponding 
value of $\Q$ at the location of the maximum is large.  

Given this, the natural question is to determine how large $\Ninf$ might become if we
push this process still further.   
Unfortunately, due to numerical limitations,  
full contour plots of the sort shown in Fig.~\ref{fig:Ninf} become increasingly difficult
to obtain
as $\lambdabar_0$ is taken increasingly small
(or as ${\cal Q}$ is taken increasingly large).
However, now that we have identified the location of parameter space in which $\Ninf$ is maximized,
and given that this region is largely independent of $\lambdabar_0$, 
we can focus our numerical analysis to this smaller relevant region
in order to study how
the maximum value of $\Ninf$ varies as a function of~${\cal Q}$.

\begin{figure}[t]
    \begin{center}
    \includegraphics[keepaspectratio, width=0.49\textwidth]{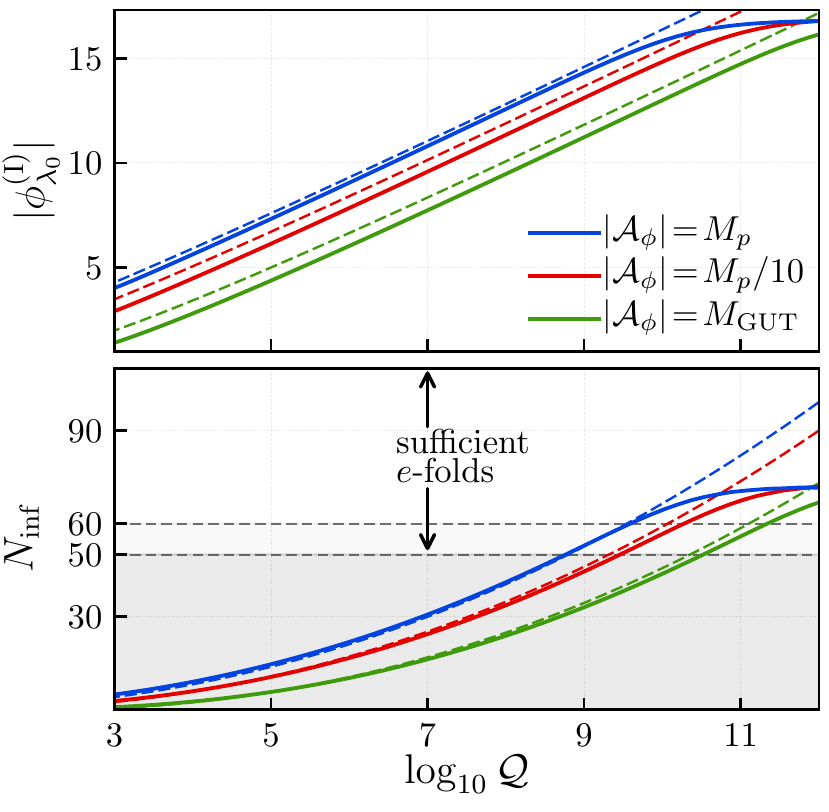}
    \end{center}
    \caption{The maximum field VEV $|\phiLNI|$ reached by our slingshot prior to the onset
    of inflation (top panel) 
    as well as the the corresponding number $\Ninf$
    of $e$-folds of inflation produced (lower panel), plotted  
    as functions of $\Q$ 
    for several different values of ${\cal A}_\phi$ assuming the primary ({\il{n=1}}) resonance.
    The solid curves  of a given color indicate the exact numerical values of $|\phiLNI|$ and $\Ninf$
    that result from numerically solving the evolution equations
    in Eqs.~\eqref{eq:eqnsofmotion} and~\eqref{eq:sourceeq} in our preferred
    region of parameter space with {\il{\mbar_{\rm sum}^2=M^2}} and {\il{\alphabar=0.9}}, while
    the dashed curves indicate the results of analytical approximations to be
    discussed in Sect.~\ref{sec:Mapping}.~
    In each case, we conclude that for sufficiently large ${\cal Q}$, our slingshot mechanism is indeed capable
    of yielding inflationary epochs of sufficient duration to solve
    the horizon and flatness problems 
    --- even when the inflaton starts with initial conditions
    from which inflation would not otherwise have been possible.} 
\label{fig:Ninf_vs_Q}
\end{figure}

Our results are shown in Fig.~\ref{fig:Ninf_vs_Q}, where
we plot 
the maximum field VEV $|\phiLNI|$ reached by our slingshot (top panel) 
as well as the 
corresponding number $\Ninf$
of $e$-folds of inflation produced 
as functions of $\Q$ 
for several different initial conditions parametrized by ${\cal A}_\phi$ (solid colored curves).
For this figure we have taken {\il{M=M_p/3}}, {\il{\mbar_{\rm sum}^2=M^2}}, {\il{\alphabar=0.9}}, 
and the primary resonance {\il{\Delta_G=\Delta_G^{(1)}}}.
In all cases, we see that the maximum value of $\Ninf$ can easily exceed
our target range {\il{N_{\rm inf}\sim {\cal O}(50-60)}}, provided ${\cal Q}$ is
sufficiently large.   
Indeed, with these parameter choices 
the precise value {\il{\Ninf=60}} corresponds to {\il{Q\approx 4\times 10^{9}}} 
for {\il{|{\cal A}_\phi| = M_p}}, or equivalently {\il{\lambdabar_0\approx 10^{-11}M_p}}. 
{\it Thus, we conclude that our slingshot mechanism is indeed capable
of yielding inflationary epochs of sufficient duration to solve
the horizon and flatness problems --- even when the inflaton starts with initial conditions
from which inflation would not otherwise have been possible.} 

Interestingly, we also see from Fig.~\ref{fig:Ninf_vs_Q} 
that $\Ninf$ does not grow without bound as ${\cal Q}$ is
increased, but eventually saturates somewhat beyond our target range.
This is apparently a direct result of the dynamics of this system.
However, there is also another limitation on the size of $\Ninf$.
Increasing the value of $\Ninf$ corresponds to increasing the abruptness
with which our scalar-field eigenvalues $\lambda_{0,1}$ (and the mixings 
between the corresponding eigenstates)
vary during the
phase transition.  Such changes render the evolution of the potential 
non-adiabatic, which in turn    results in the production of particles 
and a concomitant loss of energy density from the zero-modes of
$\phiLN$ and $\phi_{\lambda_1}$.   This too can ultimately suppress the maximum
attainable value of $N_{\rm inf}$.  This particle-production effect 
and the corresponding upper bound it implies for $\Ninf$ are discussed 
in greater detail in Appendix~\ref{sec:ParticleProduction}.~   The upshot,
however, is that our main conclusion is unchanged.  Consequently our slingshot
mechanism can indeed yield inflationary epochs of sufficient duration, as described above.

\begin{figure}[t]
    \begin{center}
        \includegraphics[keepaspectratio, width=0.499\textwidth]{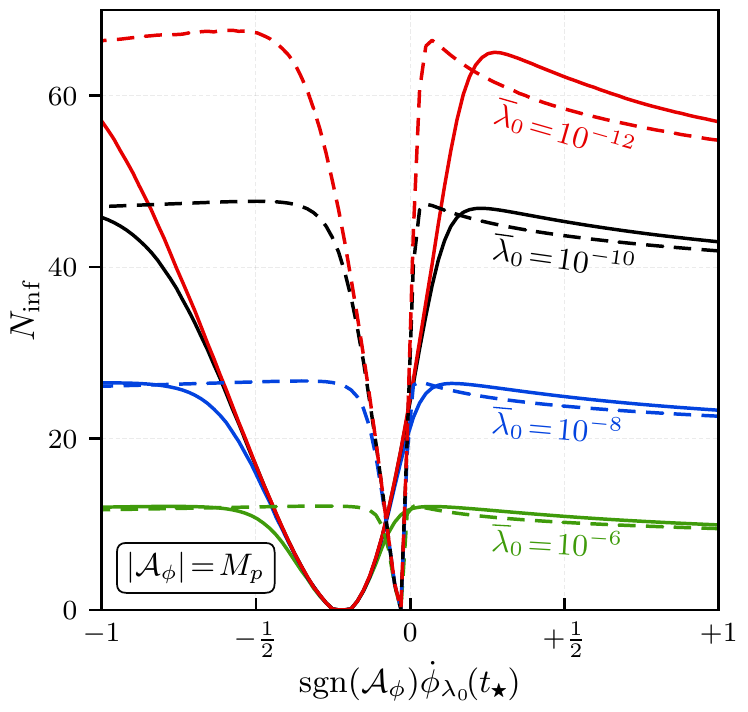}
    \end{center}
    \caption{The number of inflationary $e$-folds produced by our slingshot mechanism, 
    plotted as a function of the initial field velocity $\phidot_0(t_{\medwhitestarcap})$
    prior to the phase transition.
    The solid and dashed lines correspond to slingshots with
      {\il{\rho_R(t_\medwhitestarcap)= 0.1\, M_p^4}} and {\il{\rho_R(t_\medwhitestarcap)= 10^{-3} M_p^4}}, respectively.
    Despite the emergence of a strong localized suppression 
    when ${\cal A}_\phi$ and $\phidot_0(t_{\medwhitestarcap})$
    are of opposite signs, we see that $\Ninf$ remains large and approximately constant
    when ${\cal A}_\phi$ and $\phidot_0(t_{\medwhitestarcap})$ are of the same sign.
    Thus our results are fairly insensitive to the precise value of
    $\dot\phi_0 (t_\medwhitestarcap)$ in such cases, 
     and no fine-tuning of this variable is needed.}
\label{fig:phidot}
\end{figure}

As we discussed in Sect.~\ref{sec:TheScalarSector}, 
our slingshot mechanism has thus far been built on the assumption
that {\il{\phi_0(t_{\medwhitestar})={\cal A}_\phi}} but that {\il{\dot \phi_0(t_{\medwhitestar})=0}}. 
A natural question, then, is to examine what might happen if we loosen this last restriction
and consider arbitrary values of $\dot\phi_0(t_{\medwhitestar})$.
In Fig.~\ref{fig:phidot} we plot the resulting values of $\Ninf$ as a function
of $\dot\phi_0(t_{\medwhitestar})$ for a variety of different masses $\lambdabar_0$.
In this plot the solid and dashed lines respectively correspond to primary-resonance slingshots with
{\il{\rho_R(t_\medwhitestar)= 0.1\, M_p^4}} and 
{\il{\rho_R(t_\medwhitestar)= 10^{-3} M_p^4}},
and we have taken {\il{{\cal A}_\phi= M_p}} along with our usual benchmark
choices {\il{M=M_p/3}} and {\il{\alphabar=0.9}}.
Interestingly, when ${\cal A}_\phi$ and $\phidot_0(t_\medwhitestar)$ 
are of opposite signs, our field is first propelled towards {\it smaller}\/ field VEVs
before ultimately experiencing the slingshot.
This then results in a marked suppression in the total number of inflationary $e$-folds produced.
Indeed, we see that the behavior of $\Ninf$ in the highly-suppressed region 
is largely independent of the mass $\lambdabar_0$ of the inflaton.
However, we also see from 
Fig.~\ref{fig:phidot} that we can nevertheless achieve large values of $\Ninf$ even 
when {\il{\dot\phi_0 (t_\medwhitestar) \not= 0}}.
Indeed, when 
${\cal A}_\phi$ and $\dot\phi_0 (t_\medwhitestar)$ are of the same sign,
our results for $\Ninf$ are approximately independent of 
$\dot\phi_0 (t_\medwhitestar)$, as claimed in Sect.~\ref{sec:TheScalarSector}.
Thus we do not need to fine-tune the value of 
$\dot\phi_0 (t_\medwhitestar)$ in order to achieve our results.

\begin{figure}[t]
    \begin{center}
        \includegraphics[keepaspectratio, width=0.499\textwidth]{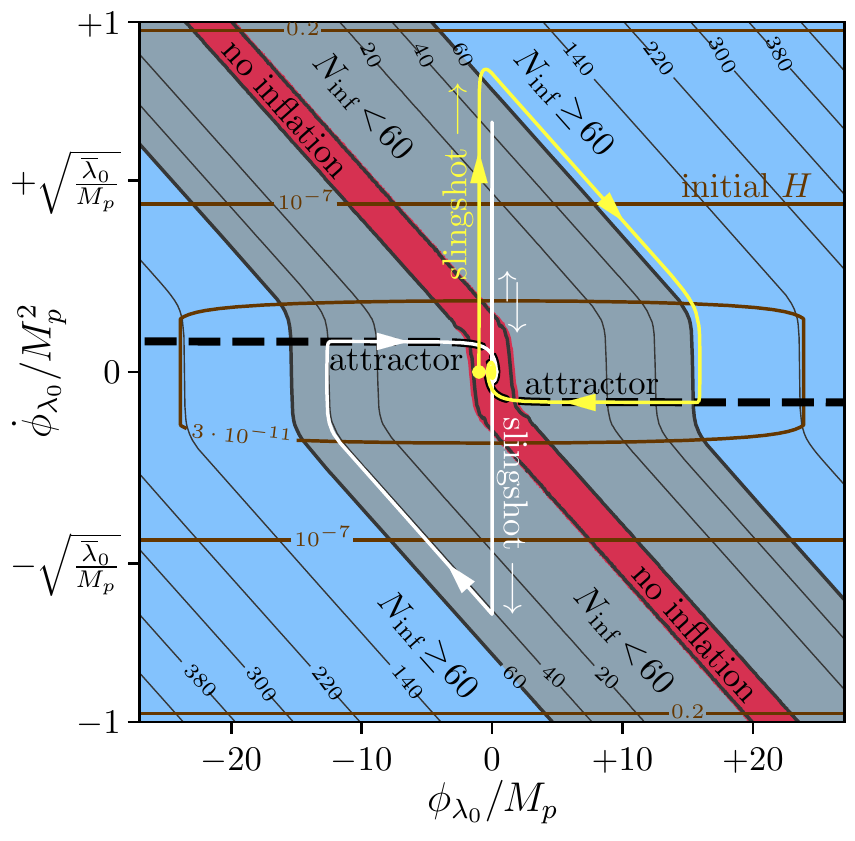}
    \end{center}
    \caption{
    The trajectory of the inflaton field $\phi_{\lambda_0}$ within the 
    \il{\{\phiLN, \phidot_{\lambda_0}\}}
    phase space for different slingshots: the \il{n=1} resonance with 
    \il{\Aphi= -M_p} 
    (yellow), and the \il{n=2} resonance with \il{\Aphi = -M_{\rm GUT}} (white).  
    Each slingshot is plotted for {\il{\mbar_{\rm sum}^2=M^2}}, {\il{\alphabar=0.9}}, and {\il{\lambdabar_0= 3\times 10^{-12}M_p}}.
    The background regions and contours are similar to those in 
    Fig.~\ref{fig:intro_phase_space}, and are drawn for
    a quadratic inflaton potential \il{V(\phiLN) = \half \lambdabar_0^2\phiLN^2}.
    In the first case (yellow), our slingshot launches
    the inflaton from the red region upwards into the blue 
    region, producing {\il{\Ninf\approx 65}} $e$-folds of inflation, while in the second case (white) our slingshot
    ultimately launches the inflaton from the red region downwards into the gray region, producing {\il{\Ninf\approx 42}} 
    $e$-folds of inflation.  Many other slingshot trajectories are also possible.}
\label{fig:outro_phase_space}
\end{figure}

Finally, 
while 
the plots shown in Fig.~\ref{fig:phirho_dominant} illustrate the cosmological
evolution resulting from our slingshot mechanism,
it is also instructive to understand our slingshot mechanism from the perspective
of an inflaton phase-space diagram of the sort shown in Fig.~\ref{fig:intro_phase_space}.~
In Fig.~\ref{fig:outro_phase_space} we show the same phase-space diagram 
as in Fig.~\ref{fig:intro_phase_space}
assuming the same quadratic inflaton potential \il{V(\phiLN) = \half m_\phi^2 \phi^2}
where we now identify the inflaton field $\phi$ as $\phi_{\lambda_0}$ and
the inflaton mass $m_\phi$ as $\lambdabar_0$, and where we have now chosen
{\il{\lambdabar_0=3\times 10^{-12}M_p}}. 
Superimposed on this diagram we also plot two possible inflaton trajectories:
the first (yellow) illustrates the slingshot
trajectory with {\il{\Aphi= -M_p}} and {\il{\Delta_G=\Delta_G^{(1)}}} (corresponding to
{\il{{\cal Q}\approx 1.2\times 10^{10}}} and {\il{\Ninf \approx 65}},
in accordance with the results in Fig.~\ref{fig:Ninf_vs_Q}),
 while 
the second (white) shows the trajectory with 
{\il{\Aphi= -M_{\rm GUT}}} and 
{\il{\Delta_G=\Delta_G^{(2)}}} (corresponding to {\il{\Ninf\approx 42}}),
where   
{\il{M_{\rm GUT} \approx 2\times 10^{16}}}~GeV.~
In each case all other parameters are chosen as in Fig.~\ref{fig:Ninf_vs_Q}.~
The yellow and white inflaton trajectories each begin within the red region,
signifying that their initial configurations would not ordinarily lead to inflation.
However, in each case it is the slingshot mechanism which propels the inflaton
into another phase-space region from which inflation can eventually emerge:
in the first case the inflaton is propelled upwards into the blue region, 
while in the second case the inflaton is initially carried upwards but is then ultimately 
propelled downwards into the gray region --- a reversal of direction which is
consistent with the properties of the {\il{n=2}} parametric resonance,
as shown in Fig.~\ref{fig:early_fields}.

In general, for a fixed inflaton potential, we have seen in Fig.~\ref{fig:intro_phase_space} that
any inflaton trajectory is restricted to follow contour lines of constant $N_{\rm inf}$ until reaching an attractor.
It may therefore seem surprising that our slingshot is capable of propelling our inflaton field across
such contour lines into new regions of phase space which differ so fundamentally 
in their ability to generate inflation.
However, as discussed in Sect.~\ref{sec:TheScalarSector},
at the root of our slingshot mechanism is a {\it cosmological phase transition}\/ which
introduces a fundamental time-dependence into our effective inflaton potential $V_{\rm eff}(\phi_i,t)$. 
It is this time-dependence which allows our inflaton trajectories to cross such contour lines.
Phrased slightly differently,
this explicit time-dependence in the inflaton potential indicates that the phase
transition induces 
a flow of energy into/out of  the scalar sector,
and it is this flow of energy which can kick the inflaton field to new locations in phase space.
Additionally, this time-dependence can endow the heavier scalar $\phi_{\lambda_1}$ with a non-negligible portion of 
the energy density, so that the energy density of the universe is not entirely dominated 
by that of the lighter field.  
However, after the phase transition has ultimately passed 
and the energy density of the heavier field has been diluted to negligible levels,  
our model is effectively described by a single light field $\phiLN$ moving 
in a quadratic potential
\beq
    V(\phiLN) ~=~ \half \lambdabar_0^2 \phiLN^2  ~,
\label{eq:quadratic}
\eeq
with $\phiLN$ starting at the new phase-space location to which it has been propelled by the slingshot.
Indeed, from this point forward, 
the inflaton will then follow the usual contour lines.
Thus, it is only during the phase transition that the slingshot can work its magic,
kicking the inflaton into a new region of phase space from which the standard inflationary
dynamics then operates in order to produce inflation.

\FloatBarrier
\section{Mapping to inflationary observables\label{sec:Mapping}}


Having demonstrated the success of our slingshot mechanism in producing inflationary epochs of
sufficient duration, we now turn to a somewhat more theoretical issue. 
Normally one would parametrize an inflationary model in terms of the VEV $\phiLNI$ 
of the inflaton at the moment when inflation commences.
Indeed, 
the value of $\phiLNI$ in turn 
determines most of the quantities of interest which characterize the inflationary 
epoch, including $\Ninf$.  
However, our slingshot mechanism and its associated phase transition
furnish us with dynamics that occurs {\it prior}\/ to this point.
As such, this mechanism effectively establishes a map from parameters 
that characterize the phase transition to the inflaton VEV itself:
\beq
    \{M, \mbar_{ij}, \Aphi, \Delta_G, t_G\} ~\longrightarrow~ \phiLNI ~ .
\label{eq:mapping}
\eeq
In this section, we shall construct this map analytically for the
region of parameter space where $\Ninf$ is a maximum, and then use this tool to elucidate
different behaviors of the slingshot mechanism.
These results will also be useful when we discuss several constraints on our
mechanism  in Sect.~\ref{sec:Constraints}.~
For concreteness, we shall focus on the case in which {\il{n = 1}}, 
but we emphasize that qualitatively similar results are obtained 
for higher-order resonances as well.

In order to derive our analytic expression for $\phiLNI$, we begin by noting that  
$\phiLN$ undergoes a period of underdamped oscillation during the phase transition, 
prior to the onset of re-overdamping.  Throughout this period, $\phiLN$ oscillates with a 
frequency approximately equal to the peak mass \il{\lambda_0(\tp)} which the inflaton 
attains as a result of the pulse.  The velocity of the field as it is released by the 
slingshot is therefore approximately \il{\phidot_{\lambda_0} \sim \Aphi\lambda_0(\tp)}. 
Within our region of interest, the corresponding kinetic-energy density 
\il{\half \phidot_{\lambda_0}^2 \sim \half \Aphi^2\lambda_0^2(\tp)} is 
sufficiently large that the universe enters a brief epoch of kination-domination immediately
after this release, wherein \il{w\approx \wphi \approx 1}.  

During this epoch, however, the kinetic-energy density with which $\phiLN$ is 
initially endowed is rapidly dissipated by Hubble damping, 
falling as {\il{\rho_\phi\sim a^{-3(1+w_\phi)} \sim a^{-6}}}  until essentially 
only potential energy remains.  
We can therefore determine the number of \e-folds of kination-domination $\Nkin$ 
by taking the ratio of $\rho_{\phi}$ between the beginning 
and ending of this period:
\beq
  e^{6\Nkin} ~\approx~ 
    \frac{\half [\lambda_0(\tp)\Aphi]^2}{\half [\lambdabar_0\phiLNI]^2}
    ~\approx~ \Q^2 \frac{\Aphi^2}{[\phiLNI]^2} \ .
  \label{eq:Nkinreduction}
\eeq
Here $\Q$ is the mass ratio defined in Eq.~\eqref{eq:Qdef}.
Meanwhile, during the kination-dominated epoch, $\phiLN$ evolves from approximately its initial value
{\il{\phiLN \approx \mathcal{A}_\phi}} to its value $\phiLNI$ 
when inflation begins.
Since the rate of change in $\phiLN$ 
during the
kination-dominated epoch is given by Eq.~\eqref{eq:dphidNkin}, and since we are focusing
our attention on the {\il{n=1}} resonance, we have 
\beq
  \phiLNI ~\approx~  \Aphi - {\rm sgn}(\Aphi) \, \sqrt{6}M_p\Nkin ~.
  \label{eq:Nkinfall}
\eeq

By eliminating $\Nkin$ between Eqs.~(\ref{eq:Nkinreduction}) and~(\ref{eq:Nkinfall}), 
we may obtain a rough estimate for the value $\phiLNI$.  However, in order to refine this 
estimate, we incorporate a correction factor into Eq.~(\ref{eq:Nkinreduction}) which
compensates for the fact that this relation was derived within the approximation that the 
transitions into and out of the kination-dominated epoch are effectively instantaneous.  
This correction factor may be parametrized as a linear shift in $\Nkin$ of the form 
\il{\Nkin \rightarrow \Nkin + \delta\Nkin}.  By comparing the results obtained from 
Eq.~(\ref{eq:Nkinreduction}) with our numerical results, we find that 
\il{\delta\Nkin \approx -1/3}.  Incorporating this correction, we find that 
$\phiLNI$ is approximately 
\beq
  \big|\phiLNI\big| ~\approx~ 
  \sqrt{\frac{2}{3}}\, M_p \, W\!\!\left(\frac{2\sqrt{3}\Q}
  {e^{\sqrt{\frac{3}{2}}\frac{|\Aphi|}{M_p}}}\frac{|\Aphi|}{M_p}\right) \ ,
  \label{eq:analyticalresult}
\eeq
where $W(x)$ denotes the Lambert $W$-function and where we have approximated 
{\il{2^{-3/2} e \approx 1}} in the argument of this function.

Remarkably, despite the complicated dynamics associated with the phase transition,
Eq.~\eqref{eq:analyticalresult} indicates that the initial value $\phiLNI$ for the 
inflaton field at the onset of inflation is principally determined by the value of $\Q$.  
In this sense, $\Q$ acts as a figure 
of merit which determines the efficacy of the slingshot mechanism.  Moreover, given
that $\Q$ is simply the ratio of the peak mass $\lambda_0(\tp)$ to the late-time
inflaton mass $\lambdabar_0$, there exists a straightforward mapping from the 
parameters $\{M, \Aphi, \mbar_{ij}, \Delta_G, t_G\}$ to $\Q$, and hence to $\phiLNI$.  
For example, within our region of interest, we have \il{\betabar \ll 1}.  
Within this regime, the late-time inflaton mass is given approximately by
\beq
    \lambdabar_0^2 ~\approx~ \half \mbar_{\rm sum}^2
      \left(1-\alphabar^2\right)\betabar ~ .
\eeq
Likewise, we find that the mass $\lambda_0(\tp)$ at the peak of the pulse is given by
\beq
\lambda_0^2(\tp) ~\approx~ 
\frac{\overline{\mathcal{M}}^2_{00}\left[1 + \frac{\mbar_{\rm sum}^2}{M^2} - 2\sqrt{\frac{\overline{\mathcal{M}}^2_{11}}{M^2}}\right]}{1 + \frac{\mbar^2_{\rm sum}}{M^2}\Big(\frac{\mbar^2_{\rm sum}}{M^2} - 2\alphabar\Big)}~.
\eeq
Since {\il{\lambdabar_0 \propto \betabar^{1/2}}} in this regime, while $\lambda_0(\tp)$ is 
independent of $\betabar$, we have the freedom to adjust these two mass scales independently 
of one another by varying the mixing saturation.  Thus, we can obtain a large $\Q$ --- 
and thus, by extension, a large field VEV $\phiLNI$ --- by 
taking $\betabar$ to be small and thereby arranging a separation between the mass scales 
$\lambda_0(\tp)$ and $\lambdabar_0$.   

From the relationship in Eq.~\eqref{eq:analyticalresult} between our 
model parameters and $\phiLNI$,
it is straightforward to determine the number of \e-folds of inflation generated
by our slingshot mechanism.  In particular, since the final value of $\phiLN$ 
at the end of inflation is \il{\phiLNE = \sqrt{2}M_p}, we find that 
\beq\label{eq:Ninf}
  \Ninf ~\approx~ 
  \int_{\phiLNIexp}^{\phiLNEexp} \!\! d\phiLN \, \frac{H}{\phidot_{\lambda_0}}
  ~\approx~ \frac{1}{4}\frac{[\phiLNI]^2}{M_p^2} - \frac{1}{2} ~,~~~
\eeq
with $\phiLNI$ given by Eq.~\eqref{eq:analyticalresult}.  We may simplify the resulting
expression by noting that {\il{\Q \gg M_p/|\Aphi| \gtrsim 1}} within our 
region of interest and invoking the approximation {\il{W(x) \approx \log(x/\log x)}} for {\il{x \gg 1}}.  
We then find that
\beq
  \Ninf ~\approx~ \frac{1}{6}\log^2
  \left[\!\frac{2|\Aphi|\Q\,e^{\sqrt{\frac{3}{2}}\big(\!1-\frac{|\Aphi|}{M_p}\!\big)}}
  {1 + \frac{3}{\sqrt{2}}\Big(\!1\!-\!\frac{|\Aphi|}{M_p}\!\Big) + 
  \sqrt{3}\log\!\Big(\!\frac{|\Aphi|}{M_p}\Q\!\Big)}\!\right]~.~~
  \label{eq:NinfInTermsOfQ}
\eeq

In Fig.~\ref{fig:Ninf_vs_Q}, we have already plotted the exact values of
$|\phiLNI|$ and $N_{\rm inf}$ as functions of ${\cal Q}$ 
(solid curves)
using results obtained through full numerical calculations.
We now superimpose on these plots the results of the analytic 
approximations in Eqs.~(\ref{eq:analyticalresult}) and~(\ref{eq:NinfInTermsOfQ}) 
for $|\phiLNI|$ and $\Ninf$ respectively (dashed curves).
As we see from 
this figure, these two sets of results coincide well for all ${\cal Q}\lsim 10^{10}$. 
We thus conclude that Eqs.~(\ref{eq:analyticalresult}) and~(\ref{eq:NinfInTermsOfQ}) 
provide excellent approximations 
for $|\phiLNI|$ and $\Ninf$ as functions of $\Q$ for all ${\cal Q}$ within this range.
Most importantly, this includes the target range for $\Ninf$ in which we are most interested, namely 
that for which {\il{\Ninf\sim {\cal O}(50-60)}}.
Of course, these approximations fail to capture the 
suppression for $|\phiLNI|$ and $\Ninf$ that develops for ${\cal Q}\gsim 10^{10}$.
This is not unexpected, as this suppression involves 
effects beyond those involved in this analytical approximation. 

\section{Constraints\label{sec:Constraints}}


At this point we have realized our main objectives in this paper:  we have developed
a ``slingshot'' mechanism, we have shown that it can lead to an inflationary epoch, and 
we have further shown that there exist regions of parameter space wherein a sufficient 
number of \e-folds arises.  In this final section, we shall take a
cursory look beyond these primary objectives and briefly discuss the observational
consequences of this particular realization of the slingshot mechanism. 

Generally speaking, most of the useful information we have about the inflationary epoch 
is derived from the spectrum of anisotropies detected in the CMB.~  Since 
\il{\lambdabar_0 \ll H} during the inflationary epoch, our inflaton field $\phiLN$ 
experiences quantum fluctuations
\beq
  \delta \phi_{\lambda_0} ~\approx~ \frac{H}{2\pi}
\eeq
throughout this epoch.  These fluctuations are systematically ``frozen-in'' as 
classical curvature perturbations while the universe inflates.  Specifically, a 
fluctuation with comoving wavenumber $k$ effectively becomes classical once 
\il{k < aH} and the corresponding wavelength exceeds the comoving horizon.
To a zeroth-order approximation these primordial perturbations are scale-invariant,
since $H$ is approximately constant during inflation.  It is therefore sensible
to parametrize the scalar power spectrum $\mathcal{P}_{s}(k)$ and tensor power 
spectrum $\mathcal{P}_{T}(k)$ in terms of their deviations from a purely scale-invariant 
form~\cite{rubakovinflation}.  In particular, these primordial power spectra are typically 
parametrized as
\begin{eqnarray}
    \mathcal{P}_{s}(k) &=& A_{s} \left(\frac{k}{k_*}\right)^{n_s - 1} \nonumber \\
    \mathcal{P}_{T}(k) &=& A_T \left(\frac{k}{k_*}\right)^{n_T}~,
    \label{eq:scalarpowerspectrum}
\end{eqnarray}
where $A_{s}$ and $A_T$ are the respective amplitudes and
\il{n_s,n_T} are the scalar and tensor spectral indices.  The fiducial wavenumber $k_*$ which 
appears in both $\mathcal{P}_{s}(k)$ and $\mathcal{P}_{T}(k)$ corresponds to the scale of 
perturbations which exit the horizon some $\Ninf^*$ \e-folds before the end of inflation.
Since these modes leave detectable imprints in the CMB, observational data
constrains the values of the parameters $A_s$, $A_T$, $n_s$, and $n_T$ and combinations thereof.

While the slingshot dynamics which leads to the inflationary epoch
is fairly complicated, the corresponding 
predictions for inflationary observables are relatively straightforward
if $\Ninf\gg \Ninf^\ast$.
The reason is that once our phase transition has concluded, 
the inflaton potential in this model is static and purely quadratic in $\phiLN$.
For a potential of this form, most inflationary observables are effectively determined by the 
value of $\Ninf^*$ alone, provided that \il{\Ninf \gg \Ninf^*}.  
For example, the ratio of scalar-to-tensor power is given by
\beq
  r ~\equiv~ A_{s}/A_T ~\approx~ \frac{16}{1+2\Ninf^*} \ ,
\eeq
while the scalar spectral index is given by 
\beq
   n_s ~\approx~ 1 - \frac{2}{1 + 2\Ninf^*} \ ,
\eeq
and so forth.  In other words, the observational constraints which apply to our model are 
essentially the same as those which apply to other models of inflation in which the inflaton 
potential is purely quadratic, regardless of the complicated dynamics through which the
initial conditions for the inflaton field are established. 

By contrast, the amplitude $A_s$ in Eq.~\eqref{eq:scalarpowerspectrum} depends not solely on 
$\Ninf^*$, but on other model parameters as well.  In particular, 
$A_s$ is in determined by the energy scale of inflation --- or equivalently, given our potential, 
the late-time mass $\lambdabar_0$ of the inflaton --- through the relation
\begin{align}\label{eq:amplitude}
    A_{s} ~&=~ \frac{1}{12\pi^2M_p^6}\left.
      \left[\frac{V^{3/2}(\phiLN)}{V'(\phiLN)}\right]^2\right|_{\phiLN\!=\phiLN^*} \nn \\ 
    ~&=~ \frac{1}{6}\left(\frac{\lambdabar_0 [\phiLN^*]^2}{4\pi M_p^3}\right)^2 \ , 
\end{align}
where $\phiLN^*$ is the inflaton VEV at the time at which the mode with wavenumber $k_*$ 
exits the horizon.  Recent measurements by the Planck Collaboration
at the scale \il{k_*=0.05\,\Mpc^{-1}} yield an amplitude 
\il{A_{s} \approx 2\times 10^{-10}}~\cite{Akrami:2018odb} for scalar perturbations.  
Thus, in order to produce both sufficient scalar power and a sufficient amount of 
inflationary expansion, the inflaton mass must reside around the
scale \il{\lambdabar_0 = \mathcal{O}(10^{-6}M_p)} in such a scenario.

If we assume that the dominant primordial perturbations are generated by quantum fluctuations
of the inflaton field, then a tension clearly exists between Eq.~\eqref{eq:amplitude}
and other predictions which follow from this realization of our slingshot mechanism.
Of course, a significant feature of this mechanism is that the initial 
inflation-triggering 
field configuration $\phiLNI$ is not put in by hand, but rather determined
by the pre-inflationary phase transition.  Indeed, our slingshot mechanism provides a linkage 
between $\lambdabar_0$ and the field VEV $\phiLNI$ at the beginning of inflation, such that 
a degree of freedom is mapped from the inflation model to the phase transition.

This tension can be addressed in several ways.  
For example, one might consider a generalization of our minimal model in which the 
contributions to $V_{\rm eff}(\phi,t)$ from the phase 
transition include not only quadratic terms, but higher-order terms as well.
Alternatively, the primordial perturbations may 
originate from another sector, such as from the dynamics of a
curvaton field $\chi$~\mbox{~\cite{Moroi:2001ct,Enqvist:2001zp,Lyth:2001nq}},
which couples negligibly 
to the scalars involved in our slingshot mechanism.
Indeed, curvatons are a standard ingredient in many models of early-universe dynamics, 
and in our case 
the degrees of freedom associated with such a curvaton sector allow us to 
relieve this tension and thereby render our slingshot mechanism consistent with the observed
perturbation spectrum.
Although at early times the curvaton would be sub-dominant \il{\rho_{\chi}\ll \rho_{\phi}}
and thereby act as a spectator to the inflationary dynamics, the curvaton is still subject
to fluctuations \il{\delta\chi \approx H/(2\pi)}, provided it 
is sufficiently light.  After inflation ends, the curvaton grows closer to dominating 
the energy density and eventually decays, transferring its
perturbations to the radiation bath~\cite{Mollerach:1989hu,Linde:1996gt}.  
The properties of the resulting perturbation spectrum
are then set not only by the Hubble scale during inflation,
but also by details of the curvaton evolution.  
Furthermore, we note that while the curvaton dynamics
depends on the inflation model, 
the slingshot mechanism we have discussed in this paper
is not affected by the introduction of the curvaton.
In general, the curvaton could
conceivably have many different types of potentials which lead to
very different field evolutions.
Of course, for the (quadratic) inflaton potential discussed in this paper, 
concave-up curvaton potentials would fall short phenomenologically since
they cannot produce a sufficiently red-tilted spectrum. 
However, as explicitly shown in Ref.~\cite{Kawasaki:2012gg}
and then argued more generally in Ref.~\cite{Kobayashi:2013bna}, 
potentials with a concave-down region can be 
phenomenologically viable.  Thus, our slingshot mechanism is not only capable 
of producing an inflationary epoch with a sufficient number of $e$-folds
but can also co-exist with a curvaton sector that helps to produce a 
phenomenologically acceptable perturbation spectrum.
 
We emphasize that the results of this section rest on the assumption that
\il{\Ninf \gg \Ninf^*}.  If this assumption does not hold, the analysis of the perturbation
spectrum becomes more subtle.  Indeed, in this case $\mathcal{P}_{s}(k)$ and 
$\mathcal{P}_{T}(k)$ ultimately depend on the state of the universe prior to inflation.  
Such considerations shall be discussed in greater detail in Sect.~\ref{sec:Conclusions}.

\section{Conclusions and discussion\label{sec:Conclusions}}


In this paper, we have proposed a novel ``slingshot'' mechanism which can enlarge the
space of viable initial field configurations for inflation.
We have illustrated this mechanism within the context of a minimal model whose core is a
sector consisting of two scalar fields undergoing a mass-generating phase transition.
However, while this minimal model contains all of the essential
ingredients necessary for our slingshot mechanism and yields a sufficient number of
$e$-folds of inflation, there are a number of possible generalizations and extensions
which merit further study.

One such possible generalization concerns
the form of our effective potential 
$V_{\rm eff}(\phi_i,t)$.  For simplicity, we have focused in this paper on the case in
which $V_{\rm eff}(\phi_i,t)$ is quadratic in the fields of the scalar 
sector.  However, in the most general case,
higher-order terms may also arise in $V_{\rm eff}(\phi_i,t)$.  The presence
of such anharmonic terms in the potential can have a significant impact on
the scalar potential at large distances in field space away from its minimum,
potentially giving rise to plateau regions or other features which could
potentially alter the inflationary dynamics and impact inflationary observables.

One natural way in which such higher-order terms in the scalar potential might be
generated is through a non-minimal coupling of the fields in the scalar sector to gravity.
In particular, one could consider scenarios in which a
non-minimal coupling between the $\phi_i$ and the scalar curvature
$\mathcal{R}$ --- \eg, a term \il{\propto \mathcal{R}\phi_i^2} --- is present
in the Jordan frame~\cite{Futamase:1987ua}.  In such scenarios, the resulting modification
of the scalar potential in the Einstein frame can help to ease tensions with
applicable phenomenological constraints.  Indeed, even in the simplest case in which
the scalar potential is quadratic in the Jordan frame, a non-minimal coupling between the
inflaton and $\mathcal{R}$ can deform the potential such that it becomes concave
down at large field values, thereby rendering such a potential viable for realistic
inflation~{\mbox{\cite{Boubekeur:2015xza,Tenkanen:2017jih}}}.  It would therefore be interesting to
investigate whether the introduction of such non-minimal couplings between the $\phi_i$ and
$\mathcal{R}$ in the context of our slingshot mechanism would have similar advantages,
enhancing the number of \e-folds, while at the same time producing a sufficient amount
of power in scalar perturbations.

An interesting feature of our slingshot mechanism is that the maximum
number of \e-folds of inflation
obtained is only slightly above the threshold value
\il{\Ninf\approx {\cal O}(50-60)} needed in order to solve the horizon
and flatness problems.
In most inflationary scenarios,
the epoch of inflationary expansion typically begins long before currently observable modes
in the perturbation spectrum exit the horizon.  In such models, the Bunch-Davies
vacuum~{\mbox{\cite{Chernikov:1968zm,Bunch:1978yq}}} provides a natural set of initial conditions for
these perturbations.
By contrast, in scenarios with \il{\Ninf\approx {\cal O}(50-60)}, the initial conditions
which determine the spectrum of large-scale perturbations differ from those associated with
the Bunch-Davies vacuum.  In particular, they  depend on the state of the universe \emph{prior} to the inflationary
epoch~{\mbox{\cite{Ramirez:2011kk,Ramirez:2012gt}}}.  In the context of our slingshot mechanism,
this spectrum in principle depends
on the state of the universe both during the phase transition itself and during the
kination-dominated phase that follows it.
Thus, in scenarios which make use of our slingshot mechanism, the predictions for a number of
inflationary observables could deviate from those associated with more standard inflationary
scenarios.
It would be interesting to investigate the
possible perturbation spectra that can arise as a consequence --- 
especially since the spectra associated with non-standard initial conditions
can exhibit certain features which are advantageous from a phenomenological perspective.
For example, many scenarios involving other kinds of modifications of the cosmological history
prior to inflation --- among them fast-roll
inflation~{\mbox{\cite{Contaldi:2003zv,Destri:2008fj,Destri:2009hn}}}, inflection-point
inflation~\cite{Downes:2012gu}, ``climbing-scalar'' scenarios in braneworld
cosmologies~\cite{Dudas:2010gi,Dudas:2012vv,Kitazawa:2014dya, 
Kitazawa:2015uda},
other inflationary scenarios in string theory~\cite{Cicoli:2013oba}, and scenarios involving
an early epoch of radiation-domination prior to
inflation~{\mbox{\cite{Powell:2006yg,Wang:2007ws}}} --- have been invoked as
explanations for the suppression of power at low multipole number in the CMB.~

Detectable cosmological imprints of our slingshot mechanism could also potentially arise
as a result of the non-adiabatic evolution of the fields within our scalar sector.
One consequence of such non-adiabaticity is, of course, particle production due to
rapid changes in the inflaton mass during the phase transition.  This is discussed in
Appendix~\ref{sec:ParticleProduction}.~  However, non-adiabaticity can have other
cosmological consequences as well.  For example, abrupt turns in the trajectory
of the inflaton in field space can not only likewise lead to particle
production, but can also alter the primordial perturbation spectrum~\cite{Konieczka:2014zja}.
Within the context of our simple two-field model,
this is not a concern:  while the mixing generated by the phase transition does induce a
turn in the inflaton trajectory, this occurs mostly prior to inflation, and
the turn is sufficiently gradual that these effects
may be safely neglected.  However, these effects could play an important role in constraining
other, more general models in which our slingshot mechanism is realized.

Along similar lines, the presence of additional heavy fields during inflation can affect
the pattern of non-Gaussianities in the CMB.  Indeed, our slingshot mechanism relies on the
mixing between the inflaton and other heavy fields in the scalar sector.  Thus, signatures of
this sort can potentially provide a way of testing our proposal.  Indeed, imprints in the CMB resulting from the
presence of such heavy fields have been studied extensively in the literature, both within the context of
quasi-single-field inflation~\cite{Chen:2009zp,Assassi:2012zq,Noumi:2012vr} and more generally via the
analysis of other, ``cosmological collider'' signatures~\cite{Arkani-Hamed:2015bza}.
That said, the extent to which the pattern of non-Gaussianities is modified by these considerations is
set primarily by the ratio \il{H/m_\Phi}, where $m_\Phi$ is the mass scale of the additional heavy field.
Hence, while the effect of these fields on the resulting pattern of non-Gaussianities will be important
for testing realizations of our mechanism with other kinds of dynamically-generated potentials --- \eg,
realizations which include heavy fields with masses closer to the Hubble scale --- the size of this effect is
typically small for the model considered in this paper since \il{H/\lambda_1 \ll 1}.

Another natural question regarding our slingshot mechanism
concerns whether it can be applied more than once.
Although we have shown that an adequate number of \e-folds
of inflation can arise from a single slingshot, it would be interesting to investigate
whether an additional, subsequent cosmological phase transition could induce a \emph{second}
slingshot, thereby propelling the lighter field in the scalar sector to even larger
VEVs.  Indeed, one might even imagine being able to obtain
\il{\Ninf\approx {\cal O}(50-60)} \e-folds of inflation --- or more --- through a {\it sequence}\/
of successive slingshots.
At first glance, it might not seem that multiple slingshots could occur in
succession, given that any radiation present in the universe prior to the onset
of any inflationary epoch would be inflated away during that epoch and
would thus no longer be present after inflation in order to trigger the next phase transition.
This is not the case, however.
As we have seen in Sect.~\ref{sec:Inflation}, the period of inflation precipitated by
our slingshot mechanism typically begins only after both the corresponding phase
transition and the kination-domination epoch which follows the slingshot have
essentially concluded.  This opens up the possibility that a second phase
transition --- and hence another slingshot --- could occur after the end of the first
phase transition but {\it before}\/ inflation begins.
Of course, this presupposes that a suitable set of initial conditions for the
second slingshot arises after the first slingshot, and so forth.
However, the state of the scalar sector at end of the kination-dominated epoch
is one in which both fields are effectively at rest,
in which a significant hierarchy exists between the masses of the two mass-eigenstates,
and in which the lighter of these two states (which is sufficiently light that it
effectively behaves as vacuum energy) is displaced from the minimum of the potential.
These are precisely the initial conditions upon which our slingshot is built.

Unfortunately, a significant amount of energy density is dissipated from the scalar
sector by Hubble friction during the kination-dominated epoch which follows
each slingshot.  For this reason, it may be challenging to obtain
a significant overall enhancement in $\phiLNI$ --- and hence in $\Ninf$ ---
through successive slingshots.  This issue is currently under study~\cite{Dienes:future1}.

On a final note, we comment on the extent to which fine-tuning is required
in order for our slingshot mechanism to yield an inflationary epoch of acceptable
duration.  As discussed in Sect.~\ref{sec:Inflation}, the parametric-resonance condition
{\il{\Delta_G \approx \Delta_G^{(n)}}} does not require a significant degree of fine-tuning.
In fact, as we have seen in Fig.~\ref{fig:dGres}, a reasonable amount of fine-tuning would
actually be required in order for $\phi_{\lambda_0}$ to {\it avoid}\/ experiencing a substantial
resonant enhancement upon its release from the slingshot.  
Similarly, we argued in Sect.~\ref{sec:Inflation} that we do not need to fine-tune 
our initial inflaton velocity $\phidot_{\lambda_0}(t_{\medwhitestar})$, provided that it 
has the correct sign.  (Indeed, in other inflation scenarios this may no longer be 
true~\cite{
    Brandenberger:2003py,
    Albrecht:1986pi,
    Goldwirth:1990iq,
    Goldwirth:1991rj,
    Brandenberger:1990wu,
    Kung:1989xz,
    Goldwirth:1989pr,
    Goldwirth:1990pm}.)  
There is, however, one fine-tuning 
which is required in order for our slingshot mechanism to give rise to
a suitable inflationary epoch.
In order to ensure that we obtain an inflationary epoch of sufficiently long duration,
our scenario requires a fairly large value of ${\cal Q}$, which in turn requires
a significant degree of mixing between the two fields of our scalar sector, with
\il{\betabar\ll 1}.  However, this is tantamount to demanding that \il{\lambdabar_0\ll H}.
Although commonly a requirement in many models of inflation, such a small inflaton mass is unnatural
unless protected by some symmetry (\eg, a global shift symmetry) and is generally difficult to
control in the presence of Planck-suppressed operators.  These issues constitute
nothing other than the well-known $\eta$-problem.
Of course, it has not been our goal in this paper to eliminate or ameliorate these fine-tuning problems;
rather, our goal has been to provide a slingshot-based method of establishing viable initial conditions
for inflation to occur.   Our slingshot mechanism can nevertheless be viewed as
{\it reformulating}\/ these fine-tuning issues in terms of the mixing structure of
a scalar sector, as established by a pre-inflationary phase transition.
This could potentially provide an alternative route toward addressing
these fine-tuning issues.

\section*{Acknowledgments}


We are happy to thank S.~Watson for many helpful discussions.
The research activities of KRD are supported in part by the U.S. Department of Energy
under Grant DE-FG02-13ER41976 (DE-SC0009913) and by the U.S. National Science Foundation
through its employee IR/D program, while those of JK were supported by the Institute
for Basic Science in Korea under project code IBS-R018-D1 
and those of BT are supported in part by the National Science
Foundation under Grant PHY-1720430.  
Portions of this work were performed at the Aspen Center for Physics,
which is supported in part by the National Science Foundation
under Grant PHY-1607611.
The opinions and conclusions expressed herein are those
of the authors, and do not represent any funding entities.

\bigskip

\appendix

\FloatBarrier
\section{Particle production in the scalar sector\label{sec:ParticleProduction}}


In this paper, we have framed our discussion of the scalar-field dynamics associated
with our slingshot mechanism largely in terms of the motion of the classical fields.  
However, certain quantum-mechanical effects can have an appreciable impact on 
the cosmology of our scalar sector and must also be taken into account.
In particular, in scenarios in which the masses of the fields change abruptly,
there can be significant production of particle-like excitations 
with comoving momenta $k$ and corresponding energies $\omega_k$ for which
the adiabaticity condition
\beq
  \abs{\frac{1}{\omega_k^2}\frac{d\omega_k}{dt}} ~\ll~ 1
  \label{eq:adiabaticity}
\eeq
is violated~{\mbox{\cite{Birrell:1982ix,Kofman:1994rk,Shtanov:1994ce,Kofman:1997yn}}}.
Bounds on the energy density associated with such particle-like 
excitations of the fields in our two-scalar model ultimately lead to an upper bound on 
$\Ninf$ for any given choice of model parameters.  Indeed, any energy density pumped into 
particle production dilutes the energy density carried by the homogeneous scalar fields, 
thereby compromising the efficacy of our slingshot mechanism. 

In this Appendix, we derive a rough estimate for the total energy density 
$\rho_{\lambda_0}^{\rm (p)}$ of 
particle-like excitations generated during the phase transition in our model 
and assess the extent to which the presence of this energy density at the onset of
inflation impacts our results.  Since the heavier mass $\lambda_1^2$ typically evolves 
adiabatically before, during, and after the phase transition, we focus solely on the 
contribution associated with the lighter field $\phiLN$, which plays the role of the 
inflaton.  

In the mass eigenbasis, the evolution of $\phi_{\lambda_0}$ is governed by the equation 
of motion~\cite{Dienes:2015bka}
\begin{align}
    \ddot{\phi}_{\lambda_0} + 3H\phidot_{\lambda_0} + 
      &\left(\lambda_0^2 - \thetadot^2\right)\phiLN \nn \\
    ~&=~ -2\thetadot\phidot_{\lambda_1} - \left(\thetaddot + 3H\thetadot\right)\phi_{\lambda_1}~,~~ 
\end{align}
where $\theta$ is the time-dependent mixing angle defined in Eq.~\eqref{eq:theta}.
The mixing-dependent terms, which are induced by the rotation of 
the mass eigenbasis during the phase transition, are negligible within regions 
of parameter space which give rise to inflation.  Dropping these terms and making the 
field redefinition \il{\varphi\equiv a^{3/2}\phiLN}, we find that
\beq
  \ddot{\varphi} + \left[\lambda_0^2 + 
    \left(\frac{3}{2}H\right)^2 w\right]\varphi ~=~ 0 ~,
    \label{eq:VarphiDDot}
\eeq
where $w$ is the equation-of-state parameter for the universe as a whole.
At times soon before or after the phase transition, \il{\lambda_0 \ll H},
and thus the second term inside the square brackets in Eq.~\eqref{eq:VarphiDDot} provides the 
dominant contribution.  However, during the phase transition 
the inflaton field becomes underdamped, with \il{3H \lesssim 2\lambda_0},
as a result of the pulse in the time-development of $\lambda_0(t)$.  
During a brief interval around {\il{t\sim t_G}}, then,
the first term in the square brackets in Eq.~\eqref{eq:VarphiDDot} dominates.
It is this interval during which the effects of non-adiabaticity are the most
pronounced.

In order to estimate the amount of particle production due to these considerations, 
we must track the occupation numbers of all the momentum modes, treating the homogeneous 
zero-mode of $\varphi$ as a classical, time-dependent background.  We begin by 
recalling that the quantum inflaton field $\varphihat$ is represented in the 
Heisenberg picture as
\beq
  \varphihat(\ve{x},t) ~=~ 
    \int\!\frac{d^3k}{(2\pi)^{3/2}} \left[\widehat{a}_{\ve{k}}\varphi_k(t)e^{-i\ve{k}\cdot\ve{x}} 
   + \widehat{a}_{\ve{k}}^{\dagger}\varphi_k^*(t)e^{+i\ve{k}\cdot\ve{x}}\right] ~,
\eeq
where $\varphi_k(t)$ is the Fourier mode of $\varphi(t)$ with comoving momentum $k$, and
where $\widehat{a}^{\dagger}_{\ve{k}}$ and $\widehat{a}_{\ve{k}}$ are the 
time-independent creation and annihilation operators for this mode. 
Each of the $\varphi_k(t)$ satisfies an equation of motion of the form
\beq
  \ddot{\varphi}_k + \omega_k^2\varphi_k ~=~ 0~,
  \label{eq:EOMforkModes}
\eeq
where the corresponding frequency
\beq
  \omega_k^2 ~=~ \frac{k^2}{a^2} + \meff^2 
  \label{eq:omegak}
\eeq
inherits a time-dependence both from the redshift of the physical momentum \il{k/a} and 
from the evolution of the effective inflaton mass $\meff$.  This effective mass, which is 
given by
\beq
  \meff^2 ~\equiv~ \lambda_0^2 + \left(\frac{3}{2}H\right)^2 w \ ,
  \label{eq:meff}
\eeq
evolves in time both as a result of the scalar-field dynamics associated with the 
phase transition and as a result of Hubble expansion.  Moreover, while the 
first term in Eq.~\eqref{eq:meff} is completely specified by Eq.~\eqref{eq:massspectrum}, 
the second depends on the state of the universe at times prior to the phase transition --- and 
in particular on the equation-of-state parameter $w$ for the universe at such times.   

\begin{figure}[t]
    \begin{center}
    \includegraphics[keepaspectratio, width=0.48\textwidth]{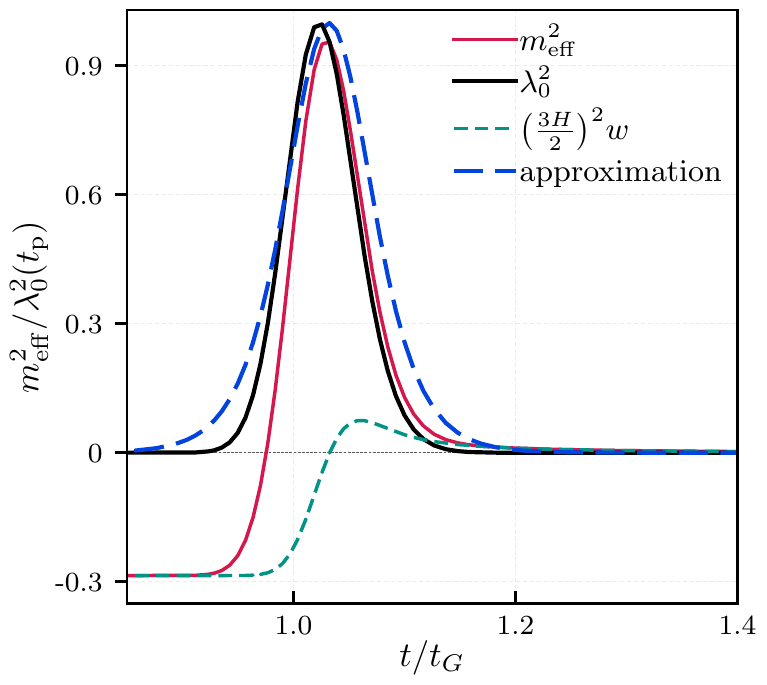}
    \end{center}
    \caption{The time-development of the effective mass $m_{\rm eff}^2$ of the 
    inflaton in our two-scalar model in the vicinity of the pulse.  The red curve
    shows the result obtained for an initial equation-of-state parameter 
    {\il{w = -1}} for the universe immediately before the phase transition. 
    The solid black and dashed green curves shown in the figure represent the 
    corresponding individual contributions from the individual terms $\lambda_0^2$ and 
    $(3H/2)^2 w$ in Eq.~\eqref{eq:meff}, respectively.  The dashed blue curve represents 
    the analytic approximation in Eq.~(\protect\ref{eq:meffapprox}).}
    \label{fig:sech2mass}
\end{figure}

In Fig.~\ref{fig:sech2mass}, we show how the value of $\meff^2$ evolves as a function of 
time during the phase transition, normalized to the value of $\lambda_0^2(\tp)$.
The red curve shows the result obtained for an initial equation-of-state parameter 
{\il{w = -1}} for the universe immediately before the phase transition.  The solid black and 
dashed green curves shown in the figure represent the contributions from the 
individual terms $\lambda_0^2$ and $(3H/2)^2 w$ in Eq.~\eqref{eq:meff}, respectively, as time
evolves.  We observe that in the vicinity of the pulse, where the change in the effective mass 
is most abrupt and the departure from adiabaticity is expected to be greatest, $\lambda_0^2$ 
represents the dominant contribution to $\meff^2$.

In order to derive an approximate analytic expression for the energy density,
we must determine the Bogoliubov coefficients which relate the asymptotic solutions 
to the equation of motion in Eq.~\eqref{eq:EOMforkModes} for the $\varphi_k(t)$ at
early times {\il{t \ll t_G}}, before the phase transition begins, to the corresponding 
asymptotic solutions at late times {\il{t \gg t_G}}, after it effectively concludes.
Of course,
the values for these coefficients depend on how the 
effective mass $\meff^2$ evolves during the phase transition. 
For simplicity, and in order to make the parametric dependence of our results more
transparent, we shall proceed by adopting a simple analytic approximation for 
$\meff^2$ --- an approximation which shall allow us to derive a corresponding 
analytic approximation for $\rho_{\lambda_0}^{\rm (p)}$.  In particular, we 
find that the function
\beq
  \meff^2 ~\approx~ \lambda_0^2(\tp)
    \,\text{sech}^2\left(\frac{t-\tp}{\frac{1}{2}\Delta_G}\right) 
  \label{eq:meffapprox}
\eeq
provides a good approximation to both $\lambda_0^2$ and $\meff^2$ 
in the vicinity of the pulse.  In fact, the blue dashed curve in Fig.~\ref{fig:sech2mass}
actually agrees well with the $\lambda_0^2$ curve {\it throughout}\/ the time interval shown.

Unfortunately, the blue dashed curve does not agree with the $\meff^2$ curve prior to the 
pulse.  However, we expect that this deviation will not have a large impact on our results.
The fact that {\il{\meff^2 < 0}} in the asymptotic regime in which {\il{t \ll t_G}} implies that the
$\varphi_k(t)$ with {\il{k \lesssim Ha}} are tachyonic, with {\il{\omega_k^2 < 0}}.  
However, such modes, whose wavelengths exceed the Hubble radius at production, cannot properly 
be considered to be particle-like excitations~\cite{Konieczka:2014zja}.  These modes, which 
behave classically, therefore do not contribute to $\rho_{\lambda_0}^{\rm (p)}$ (although
they can affect the spectrum of density perturbations).  It then follows that the spectrum of 
$\varphi_k(t)$ with higher momenta, which {\it do}\/ contribute to $\rho_{\lambda_0}^{\rm (p)}$, 
is approximately equivalent to the spectrum of modes which would be obtained had we replaced 
$\meff^2$ with $\lambda_0^2$ in Eq.~\eqref{eq:omegak}.  Given this, we shall approximate
$\meff^2$ as $\lambda_0^2$ in deriving our estimate for $\rho_{\lambda_0}^{\rm (p)}$.
We note, however, that the initial value of $w$ can have an impact on other inflationary 
observables.
   
In both the early-time and late-time asymptotic regimes, the inflaton potential evolves 
adiabatically and the $\varphi_k(t)$ are well approximated by 
\il{\varphi_k(t)\approx e^{-i\omega_kt}/\sqrt{\omega_k}}.  It can therefore be shown that
the problem of finding the Bogoliubov coefficients which relate these two sets of asymptotic
solutions for $\varphi_k(t)$ is mathematically equivalent to the problem of determining the 
transmission coefficients for scattering off a potential \il{V(x)\propto -\text{sech}^2(x)} 
in non-relativistic quantum mechanics~\cite{Amin:2015ftc}.  Thus, by analogy,
we find that the differential energy density $d\rho_{\lambda_0}^{\rm (p)}/dk$
of inflaton field quanta per unit comoving momentum $k$ at the time the 
non-adiabatic evolution effectively ceases --- which is roughly equivalent to the time
at which $\phiLN$ is released from the slingshot --- takes the form
\beq
  \frac{d\rho_{\lambda_0}^{\rm (p)}}{dk} ~\approx~  
  \frac{4\pi k^2\cos^2\big[\frac{\pi}{2}\sqrt{1 + \lambda_0^2(\tp)\Delta_G^2}\,\big]}
  {(2\pi)^3\sinh^2\left(\frac{\pi}{2}\Delta_G k\right)}~.
  \label{eq:DiffParticleQuantaBYk}
\eeq
We note that this differential energy density is largest for modes with momenta in
the regime \il{k\lesssim \Delta_G^{-1}}.  Integrating this energy density 
over $k$, we arrive at our estimate for the {\it total}\/ energy density associated with 
particle-like excitations of the inflaton at the end of the phase transition.  In
particular, at this time we find that
\beq
 \rho_{\lambda_0}^{\rm (p)} ~\approx~ \frac{6\zeta(3)}{\pi^6\Delta_G^4}
  \left[1 + \cos\left(\pi\sqrt{1 + \lambda_0^2(\tp)\Delta_G^2}\right)\right]~,
  \label{eq:rhopp}
\eeq
where $\zeta(x)$ denotes the Riemann zeta function.

In order for our slingshot mechanism to be successful, we must ensure that 
$\rho_{\lambda_0}^{\rm (p)}$ is small in comparison with the energy density 
$\rho_\lambda$ associated with the classical background fields
of the scalar sector.  At the time at which the inflaton is released from the slingshot, 
$\rho_{\lambda_0}^{\rm (p)}$ is given by Eq.~\eqref{eq:rhopp}, while  
{\il{\rho_{\lambda}\approx \frac{1}{2}[\lambda_0(\tp)\Aphi]^2}}.  Within our 
parameter-space region of interest, \il{\lambda_0^2(\tp)\Delta_G^2 \gtrsim 1} 
and the ratio of these two energy densities at this time is therefore   
\beq
  \frac{\rho_{\lambda_0}^{\rm (p)}}{\rho_{\lambda}} ~\approx~ 
  \frac{12\zeta(3)}{\pi^6\lambda_0^2(\tp)\Aphi^2}\left(\frac{1}{\Delta_G^4}\right) \ .
  \label{eq:largeL0pDG}
\eeq
During the subsequent kination-dominated epoch that precedes the onset of 
inflation, this ratio increases due to the fact that {\il{\il{\rho_{\lambda}\propto a^{-6}}}} 
during this epoch, whereas $\rho_{\lambda_0}^{\rm (p)}$, which is dominated by 
the contribution from relativistic momentum modes, scales like   
{\il{\rho_{\lambda_0}^{\rm (p)} \propto a^{-4}}}.  Thus, the ratio of these energy densities
at the end of the kination-dominated epoch, when inflation begins, is
\begin{eqnarray}
  \left.\frac{\rho_{\lambda_0}^{\rm (p)}}{\rho_{\lambda}}\right|_{\phi^{\rm ini}_{\lambda_0}}
     &\approx &~ 
    \frac{12\zeta(3)}{\pi^6\lambda_0^2(\tp)\Aphi^2}
    \left(\frac{e^{2\Nkin}}{\Delta_G^4}\right) \nonumber \\
  &\approx &~ 
    \frac{12\zeta(3) \Q^{2/3}}{\pi^6\lambda_0^2(\tp)\Delta_G^4 \big|\phiLNI\big|^{2/3} 
    \big|\Aphi\big|^{4/3}}~,~~~~
\end{eqnarray}
where in going from the first to the second line we have used 
Eq.~\eqref{eq:Nkinreduction}.  Thus, we find that for a given choice of model parameters,
the bound {\il{\rho_{\lambda_0}^{\rm (p)} \ll \rho_\lambda}} on the energy density of inflaton
field quanta at the onset of inflation ultimately translates into a bound on $\Q$ of the form
{\il{\Q \ll \Q_{\rm max}}}, where
\begin{equation}
  \Q_{\rm max} ~\equiv~ \frac{\pi^9}{24\sqrt{3}\zeta^{3/2}(3)}
    \lambda^3_0(\tp) \Delta^6_G|\phiLNI| \Aphi^2~.
  \label{eq:Qmax}
\end{equation}

It is now straightforward to assess how this bound on $\Q$ constrains our slingshot
model.
For example, for our preferred parameter choices
\il{|\Aphi| = M_p}, 
\il{\overline{m}_{\rm sum}^2 = M^2}, and \il{\alphabar = 0.9} with {\il{\Delta_G = \Delta_G^{(1)}}}, 
we find from Eq.~(\ref{eq:Qmax}) that {\il{\Q_{\rm max} \approx 3.2\times 10^{11}}}.
Comparing this upper bound on $\Q$ with the results shown in Fig.~\ref{fig:Ninf_vs_Q}, we see 
that this bound indeed imposes a non-trivial constraint on our model.  However,
we also see that our slingshot 
mechanism can nevertheless yield a sufficient number of \e-folds for cosmic inflation.
    
It is also important to note that 
while the bound on $\Q$ from particle production can be 
quite constraining, particularly for {\il{|{\cal A}_\phi| < M_p}}, there are several ways 
in which this bound can be relaxed in 
comparison to the benchmark quoted above.  First, we note that  
this bound can be considerably weaker for higher-order parametric resonances than it is for 
the primary resonance.  Indeed, since $\Delta_G^{(n)}$ increases with $n$ ---
and in fact turns out to be roughly proportional to $n$ --- 
higher-order resonances require larger values of $\Delta_G$.  Thus, given the explicit
dependence of $\Q_{\rm max}$ on $\Delta_G$ in Eq.~\eqref{eq:Qmax}, we expect   
larger values of $n$ to lead to significantly higher values of $\Q_{\rm max}$. 
While $\Q_{\rm max}$ also has an additional, implicit dependence on $\Delta_G$ through
$|\phiLNI|$ which somewhat mitigates this effect, the suppression of $|\phiLNI|$ as $\Delta_G$ 
increases is quite gradual, as indicated in Fig.~\ref{fig:dGres}.  Thus, the net impact of 
increasing $n$ is to raise $\Q_{\rm max}$ and thereby weaken the particle-production bound.  

In this same connection, we also note from Fig.~\ref{fig:dGres} that 
the parametric resonances in our slingshot model are quite 
broad.  For example, the range of $\Delta_G$ values associated with the \il{n=1} resonance 
spans several decades.  Although the corresponding range of $\Ninf$ is fairly narrow, 
the corresponding range of $\Q_{\rm max}$ is quite broad, extending over an order of 
magnitude or more.  Thus, the bounds on $\Q$ obtained for a given resonance by taking
$\Delta_G$ to be precisely equal to the corresponding $\Delta_G^{(n)}$ are likely to be overly 
conservative.

\bibliography{references}

\end{document}